\documentclass{aa}

\usepackage{graphicx}
\usepackage{txfonts}
\usepackage[colorlinks=true, linkcolor=blue, citecolor=blue, urlcolor=blue, breaklinks=true]{hyperref}
\begin{document} 

\title{Progress on the calibration of surface brightness--color relations for early- and late-type stars\thanks{based on CHARA/VEGA observations.}}

\authorrunning{Salsi et al.}

  \author{A. Salsi\inst{1}, N. Nardetto\inst{1}, D. Mourard\inst{1}, D. Graczyk\inst{2}, M. Taormina\inst{3}, O. Creevey\inst{1}, V. Hocdé\inst{1}, F. Morand\inst{1}, K. Perraut\inst{4}, G. Pietrzynski\inst{3,5}, G.H. Schaefer\inst{6}}

   \institute{Université Côte d’Azur, OCA, CNRS, Laboratoire Lagrange, France\\
              \email{anthony.salsi@oca.eu}
        \and
         Centrum Astronomiczne im. Miko{\l}aja Kopernika, Polish Academy of Sciences, Rabi\'anska 8, 87-100, Torun, Poland
         \and
         Centrum Astronomiczne im. Miko{\l}aja Kopernika PAN, Bartycka 18, 00-716, Warsaw, Poland
         \and
        Univ. Grenoble Alpes, CNRS, IPAG, 38000 Grenoble, France
        \and 
         Universidad de Concepci\'on, Departamento de Astronom\'ia, Concepci\'on, Chile
        \and
        The CHARA Array of Georgia State University, Mount Wilson Observatory, Mount Wilson, CA 91023, USA
         }         
   \date{Received... ; accepted...}

% \abstract{}{}{}{}{} 
% 5 {} token are mandatory
  \abstract
   {Surface brightness-color relations (SBCRs) are widely used for estimating angular diameters and deriving stellar properties. They are critical to derive extragalactic distances of early-type and late-type eclipsing binaries or, potentially, for extracting planetary parameters of late-type stars hosting planets. Various SBCRs have been implemented so far, but strong discrepancies in terms of precision and accuracy still exist in the literature.}
   {We aim to develop a precise SBCR for early-type B and A stars using selection criteria, based on stellar characteristics, and combined with homogeneous interferometric angular diameter measurements. We also improve SBCRs for late-type stars, in particular in the Gaia photometric band.}
   {We observed 18 early-type stars with the VEGA interferometric instrument, installed on the CHARA array. We then applied additional criteria on the photometric measurements, together with stellar characteristics diagnostics in order to build the SBCRs. 
   }
   {We calibrated a SBCR for subgiant and dwarf early-type stars. 
   %{\bf the 18 VEGA} angular diameter measurements. 
   The RMS of the relation is $\sigma_{F_{V_{0}}} = 0.0051\,$mag, leading to an average precision of 2.3\% on the estimation of angular diameters, with 3.1\% for $V-K < -0.2\,$mag and 1.8\% for $V-K > -0.2\,$mag. We found that the conversion between Johnson-$K$ and 2MASS-$K_s$ photometries is a key issue for early-type stars. Following this result, we have revisited our previous SBCRs for late-type stars by calibrating them with either converted Johnson-$K$ or 2MASS-$K_s$ photometries. We also improve the calibration of these SBCRs based on the Gaia photometry. The expected precision on the angular diameter using our  SBCRs for late-type stars ranges from 1.0\% to 2.7\%.}
   {By reaching a precision of 2.3\% on the estimation of angular diameters for early-type stars, significant progress has been made to determine extragalactic distances, such as M31 and M33 galaxies, using early-type eclipsing binaries. 
   }


   \maketitle
   
%
%-------------------------------------------------------------------

\section{Introduction}\label{Intro}

Determining the expansion of the Universe, that is the Hubble constant ($H_0$) to better than 2\%, is required in order to understand the nature of dark energy. However, the two most accurate methods for that, the cosmic microwave background \citep{2020AA...641A...6P, 2019ApJ...876...85R} and the distance ladder, are inconsistent today, which is referred to as the "tension" \citep{2020MNRAS.498.1420W}. One of the keys to resolve this tension is based on the calibration of the Leavitt period-luminosity (PL) law of Cepheids \citep{1912HarCi.173....1L}. Cepheids are indeed the backbone of the extragalactic distance ladder because their pulsation periods, which are easily determined observationally, directly correlate with their luminosities. Another method consists in using eclipsing binaries to constrain extragalactic distances.

Recently, a new estimate of the distance to the Large Magellanic Cloud (LMC), based on 20 late-type eclipsing binaries, has been obtained by the Araucaria team\footnote{\url{https://araucaria.camk.edu.pl/}} \citep{2019Natur.567..200P}.  Their precision of 1\% is mostly due to the precision of the surface brightness-color
relation (SBCR), calibrated on 41 nearby red clump giant stars using infrared interferometry \citep{2018A&A...616A..68G}. The same was done to derive the distance of the Small Magellanic Cloud (SMC) with a precision of better than 2\% \citep{Graczyk2020}. Deriving the distance from eclipsing binaries is simple: the radius of both components is estimated from the combination of photometry and spectroscopy, and angular diameters are estimated from the magnitude and color of stars through a  SBCR. The combination of radii and angular diameters provides the distance. The influence of interstellar attenuation in neighboring galaxies has been studied using several techniques so far \citep{2006ApJ...652..313B, 2019Natur.567..200P, Graczyk2020}, and it is still under investigation.

The situation concerning the early-type eclipsing binaries is more complex because the calibration of the SBCR requires high angular resolution measurements, and early-type stars are particularly active \citep{2015AA...575A..34M, 2018ApJ...869...37G}. \cite{2014AA...570A.104C} improved, by a factor of 2, the precision on SBCR of early-type stars from about 15\% to 7\%, corresponding to the most accurate SBCR developed so far for early-type stars. However, this precision is still not sufficient to derive the distance of extragalactic early-type eclipsing binaries with a precision of a few percent. It is also worthwhile to mention that the distances to M31 and M33 are currently based on models of early-type eclipsing binaries, but not SBCRs \citep{2006ApJ...652..313B, 2010AA...509A..70V}. Another interesting approach is to analyze O- and B-type detached eclipsing binaries in the LMC, for which the distance is known, and to derive the surface brightness \citep{2019ApJ...886..111T}. In this paper, we aim to calibrate the SBCR for early-type stars precisely, following the strategy of Paper I.

Moreover, with the work done on the infrared photometry of early-type stars, it is now possible to improve the SBCRs for late-type stars shown in \cite{2020AA...640A...2S} (Paper I hereafter). It turns out that some refinements are necessary concerning the infrared photometric systems that are used. Finally we have also reconsidered the calculation of the extinction in the Gaia band for proposing a new improvement of the SBCRs in this work.

Sect. \ref{data_selection}, \ref{VEGA_measurements}, \ref{SBCR_early_type} are devoted to the strategy and data selection, the VEGA observations, and the SBCR calibration of early-type stars, respectively. A subsequent discussion is provided in Sect.  \ref{discussion}. The revision of SBCR for late-type stars is presented in Sect.  \ref{improvement_SBCRs}, while some general conclusions are given in Sect. \ref{conclusion}.

\section{Strategy and data selection}\label{data_selection}

\subsection{Criteria on stellar characteristics}

In Paper I, we have shown that any stellar activity or characteristics
(multiplicity, binarity, variability, etc.) may impact the calibration
of SBCRs. We therefore implemented a set of criteria
to select a correct sample of early-type stars.

We restricted the calibration of our SBCR to $V$ and $K$ photometries, as it is the set of color which provides the lowest dispersion \citep{Kervella04b}. We started the selection from the SIMBAD Astronomical Database\footnote{Available at \url{http://simbad.u-strasbg.fr/simbad/}}. We first selected early-type stars, thus O, B, and A stars with $V-K< 1 \,$mag. We then considered only subgiants and dwarfs. From \cite{2015AA...579A.107C}, we know that the projected rotational velocity affects the surface brightness of the star. They show that a rotational velocity lower than 85\% of the critical velocity has an impact of at most $\sigma_{F_{V}} = $ 0.003$\,$mag on the RMS of the SBCR. Alternatively, if one cannot access the critical velocity of the star, they demonstrate that considering stars with projected rotational velocity $V_{\mathrm{rot}} \sin i$ lower than $100$ km/s also results in a dispersion of 0.003$\,$mag. Therefore, to reach a precision of $\sim$2\% on the angular diameter estimate, we made the choice to consider only stars with $V_{\mathrm{rot}} \sin i$ lower than 75 km/s. 

We excluded all known binary stars in our sample. The SEDs of all the stars were checked with the VO Sed Analyzer (VOSA) software\footnote{\url{http://svo2.cab.inta-csic.es/theory/vosa/}}. 

We have shown in Paper I that a variability above 0.1$\,$mag could significantly affect the SBCR. Following this strategy, we searched for information about the variability of the stars in \cite{2017ARep...61...80S}. We then rejected variable stars with a variability above 0.1$\,$mag. We quantitatively study this point later in Sect. \ref{variability}. 

We finally searched for stars with expected angular diameters between 0.3 and 0.8 milli-second of arc (mas). This is optimal for the VEGA instrument \citep{Mourard2009, Mourard2011}, installed at the CHARA array in Mount Wilson, USA \citep{CHARA}. We finally end up with a total of 18 stars to be observed in the northern hemisphere. No O-type stars were selected due to a roughly equal combination of variability and multiplicity criteria.

\subsection{Photometric selection and interstellar attenuation}\label{photo_section}

As demonstrated in Paper I, precise photometries are of course mandatory for the calibration of SBCR. 
The \citet{Kharchenko} catalog offers precise visible magnitudes and gathers measurements from several other catalogs (Hipparcos-Tycho catalogs, Carlsberg Meridian Catalog, and the Positions and Proper Motions catalog). All the visible magnitudes are given in the Johnson-$V$ filter. With such a catalog, the precision on the $V$ magnitude of our stars ranges from 0.002$\,$mag to 0.008$\,$mag.

Finding precise infrared $K$ photometry is more complex. Indeed, we faced some issues with the 2MASS catalog \citep{2MASS}, where seven out of the 18 stars in our sample are affected by imprecise infrared measurements (i.e., an uncertainty higher than 0.1$\,$mag). This problem has already been identified in Paper I and is due to saturation issues. We identified accurate measurements in \citet{Ducati} for these seven stars. 

The precision on the $K/K_s$ photometry in our sample ranges from 0.1\% to 2.7\%. However, the drawback of this approach is that the selected photometric values are inhomogeneous in terms of the filter pass band. We could convert Johnson-$K$ photometries into 2MASS using transformation equations, as we propose later in Sect. \ref{improvement_SBCRs} for late-type stars, but this leads to other issues, in particular for early-type stars, that are discussed in Sect. \ref{photometries_conv}.

For the reddening correction, we used the \textit{Stilism}\footnote{The online tool is available at \url{http://stilism.obspm.fr}} online tool \citep{Stilism1, Stilism2} to compute the color excess $E (B-V)$, considering early Gaia DR3 parallaxes \citep{2021AA...649A...1G, 2018AA...616A...1G}. The interest of this tool is the tridimensional maps of the local interstellar matter (ISM) it offers, based on measurements of starlight absorption by dust or gaseous species. The interstellar attenuation $A_{V}$ in the visible band is defined as follows

\begin{equation}
A_{V} = R_V \times E(B-V),
\end{equation}

where $R_V$ is the total-to-selective extinction ratio in the visible band, for which we set $R_V = 3.1$, and we used $A_{K} = 0.114 \times A_{V}$
%on all the $K$ photometric set 
\citep{Cardelli}. 

\section{VEGA/CHARA interferometric measurements}\label{VEGA_measurements}

\begin{table}[!htbp]
\caption{Reference stars used for VEGA observations listed with their spectral type, their $V$ magnitude, and their uniform-disk angular diameter in the $R$ band. Sect. \ref{VEGA_measurements} describes the parameters of the calibrators.}       
\label{Cals_data}   
\centering                          % used for centering table
\begin{tabular}{ccccc}        % centered columns (4 columns)
\hline\hline  
Target  &       Reference       &       Sp. Type        &       $V$     &               $\theta_{\mathrm{UD}}$[R]                                       \\
        &       stars   &               &       [mag]   &               [mas]                                   \\
\hline                                                                                                          
HD11415 &       HD10221 &       A0V     &       5.59    &       $       0.224   _{\pm   0.013   }       $       \\
        &       HD12301 &       A0I     &       5.61    &       $       0.403   _{\pm   0.013   }       $       \\
        &       HD6210  &       F6V     &       5.83    &       $       0.489   _{\pm   0.039   }       $       \\
        \hline 
HD114330        &       HD107070        &       A5IV/V  &       5.90    &       $       0.289   _{\pm   0.021   }       $       \\
        &       HD112846        &       A3III   &       5.79    &       $       0.294   _{\pm   0.009   }       $       \\
        &       HD116831        &       A8V     &       5.97    &       $       0.279   _{\pm   0.019   }       $       \\
        \hline 
HD145389        &       HD140728        &       A0V     &       5.48    &       $       0.232   _{\pm   0.015   }       $       \\
        &       HD143584        &       F0IV    &       6.03    &       $       0.327   _{\pm   0.023   }       $       \\
        &       HD144206        &       B9III   &       4.71    &       $       0.314   _{\pm   0.019   }       $       \\
        \hline 
HD145570        &       HD143459        &       A0V     &       5.53    &       $       0.272   _{\pm   0.019   }       $       \\
        &       HD145607        &       A2IV    &       5.42    &       $       0.307   _{\pm   0.023   }       $       \\
        \hline 
HD148112        &       HD144874        &       A7V     &       5.64    &       $       0.340   _{\pm   0.024   }       $       \\
        &       HD152614        &       B8V     &       4.38    &       $       0.333   _{\pm   0.024   }       $       \\
        \hline 
HD149438        &       HD146624        &       A1V     &       4.78    &       $       0.339   _{\pm   0.023   }       $       \\
        &       HD148605        &       B3V     &       4.79    &       $       0.213   _{\pm   0.015   }       $       \\
        \hline 
HD152107        &       HD143584        &       F0IV    &       6.03    &       $       0.327   _{\pm   0.023   }       $       \\
        &       HD144206        &       B9III   &       4.71    &       $       0.314   _{\pm   0.019   }       $       \\
        &       HD149303        &       A2V     &       5.68    &       $       0.286   _{\pm   0.020   }       $       \\
        &       HD155860        &       A5III   &       6.13    &       $       0.231   _{\pm   0.017   }       $       \\
        \hline 
HD192640        &       HD191610        &       B2.5V   &       4.93    &       $       0.210   _{\pm   0.018   }       $       \\
        &       HD193369        &       A2V     &       5.58    &       $       0.263   _{\pm   0.016   }       $       \\
        \hline 
HD195810        &       HD193472        &       A5      &       5.94    &       $       0.319   _{\pm   0.023   }       $       \\
        &       HD196544        &       A1IV    &       5.42    &       $       0.270   _{\pm   0.008   }       $       \\
        &       HD196740        &       B5IV    &       5.05    &       $       0.210   _{\pm   0.012   }       $       \\
        \hline 
HD27819 &       HD25202 &       F4V     &       5.87    &       $       0.376   _{\pm   0.026   }       $       \\
        &       HD28226 &       Am      &       5.71    &       $       0.345   _{\pm   0.025   }       $       \\
        \hline 
HD27962 &       HD27459 &       F0IV/V  &       5.24    &       $       0.409   _{\pm   0.029   }       $       \\
        &       HD28226 &       Am      &       5.71    &       $       0.345   _{\pm   0.025   }       $       \\
        \hline 
HD3360  &       HD1976  &       B5IV    &       5.58    &       $       0.200   _{\pm   0.006   }       $       \\
    &   HD2054  &       B9IV    &       5.72    &       $       0.207   _{\pm   0.006   }       $       \\
        &       HD3240  &       B7III   &       5.08    &       $       0.238   _{\pm   0.015   }       $       \\
        &       HD6676  &       B8V     &       5.77    &       $       0.207   _{\pm   0.014   }       $       \\
        \hline 
HD33959 &       HD34452 &       A0      &       5.37    &       $       0.214   _{\pm   0.006   }       $       \\
        &       HD34578 &       A5II    &       5.03    &       $       0.597   _{\pm   0.055   }       $       \\
        &       HD35239 &       B9III   &       5.93    &       $       0.227   _{\pm   0.016   }       $       \\
        &       HD35520 &       A1p     &       5.91    &       $       0.301   _{\pm   0.008   }       $       \\
        \hline 
HD35468 &       HD34203 &       A0V     &       5.52    &       $       0.251   _{\pm   0.016   }       $       \\
        &       HD34658 &       F3III/IV        &       5.32    &       $       0.498   _{\pm   0.038   }       $       \\
        &       HD37490 &       B3V     &       4.59    &       $       0.229   _{\pm   0.023   }       $       \\
        &       HD38899 &       B9IV    &       4.88    &       $       0.298   _{\pm   0.019   }       $       \\
        \hline 
HD58142 &       HD47100 &       B8III   &       5.33    &       $       0.236   _{\pm   0.014   }       $       \\
        &       HD56963 &       F2V     &       5.74    &       $       0.405   _{\pm   0.030   }       $       \\
        &       HD60652 &       A5m     &       5.91    &       $       0.283   _{\pm   0.019   }       $       \\
        &       HD70313 &       A3V     &       5.54    &       $       0.300   _{\pm   0.020   }       $       \\
        \hline 
HD886   &       HD1439  &       A0IV    &       5.88    &       $       0.207   _{\pm   0.013   }       $       \\
        &       HD560   &       B9V     &       5.53    &       $       0.219   _{\pm   0.013   }       $       \\
        \hline 
HD89021 &       HD85795 &       A3III   &       5.27    &       $       0.297   _{\pm   0.021   }       $       \\
        &       HD90470 &       A3V     &       6.01    &       $       0.264   _{\pm   0.007   }       $       \\
        &       HD90840 &       A4V     &       5.78    &       $       0.276   _{\pm   0.018   }       $       \\
        &       HD91312 &       A7IV    &       4.72    &       $       0.554   _{\pm   0.036   }       $       \\
        &       HD94334 &       A1V     &       4.66    &       $       0.368   _{\pm   0.011   }       $       \\
        \hline 
HD97633 &       HD92825 &       A3V     &       5.07    &       $       0.346   _{\pm   0.022   }       $       \\
        &       HD93702 &       A2V     &       5.31    &       $       0.301   _{\pm   0.022   }       $       \\
        &       HD95608 &       A1V     &       4.40    &       $       0.453   _{\pm   0.036   }       $       \\
\hline
\end{tabular}
\end{table}

\begin{table*}
\caption{VEGA angular diameter measurements for the 18 early-type stars (see Sect. \ref{VEGA_measurements} for a detailed description of the method used to derive the angular diameter of stars).} 
\label{VEGA_data}   
\centering                          % used for centering table
\begin{tabular}{ccccccccc}        % centered columns (4 columns)
\hline\hline                                                      
Name    &       Sp.Type &       $V$     &       $A_V$   &       $(V-K)_0$       &       $K$-ref &       $u_{\mathrm{R}}$        &       $\theta_{\mathrm{LD}_{\pm \sigma_{\theta_{\mathrm{LD}}}}}$        &        $\chi^2_{r}$   \\
        &               &       [mag]   &       [mag]   &       [mag]   &               &               &       [mas]   &               \\
\hline                                                                                                                                  
HD11415 &       B2V     &       3.35    &       0.050   &       $-0.484_{\pm  0.012}$        &       \cite{Ducati}   &       0.281   &       $0.471_{\pm  0.007}$        &       1.097   \\
HD114330        &       A1IV    &       4.38    &       0.012   &       $-0.022_{\pm  0.04}$ &       \cite{2MASS}    &       0.413   &       $0.443_{\pm  0.023}$    &       1.107   \\
HD145389        &       B9V     &       4.23*   &       0.028   &       $-0.115_{\pm  0.017}$        &       \cite{2MASS}    &       0.358   &       $0.452_{\pm  0.007}$        &       0.089   \\
HD145570        &       A1V     &       4.93    &       0.003   &       $0.297_{\pm  0.02}$ &       \cite{2MASS}    &       0.512   &       $0.431_{\pm  0.008}$    &       0.114   \\
HD148112        &       A2V     &       4.57*   &       0.016   &       $0.036_{\pm  0.022}$        &       \cite{Ducati}   &       0.410   &       $0.423_{\pm  0.010}$        &       0.089   \\
HD149438        &       B0V     &       2.82    &       0.214   &       $-0.939_{\pm  0.032}$        &       \cite{Ducati}   &       0.235   &       $0.338_{\pm  0.011}$        &       0.528   \\
HD152107        &       A1V     &       4.82*   &       0.022   &       $0.229_{\pm  0.022}$        &       \cite{2MASS}    &       0.434   &       $0.432_{\pm  0.009}$        &       0.134   \\
HD192640        &       A2V     &       4.95*   &       0.009   &       $0.520_{\pm  0.02}$ &       \cite{2MASS}    &       0.447   &       $0.489_{\pm  0.011}$    &       0.244   \\
HD195810        &       B6IV    &       4.03*   &       0.053   &       $-0.398_{\pm  0.038}$        &       \cite{2MASS}    &       0.319   &       $0.394_{\pm  0.003}$        &       0.038   \\
HD27819 &       A2V     &       4.80    &       0.003   &       $0.347_{\pm  0.014}$        &       \cite{Ducati}   &       0.468   &       $0.489_{\pm  0.007}$        &       1.592   \\
HD27962 &       A2IV    &       4.30*   &       0.003   &       $0.199_{\pm  0.033}$        &       \cite{2MASS}    &       0.350   &       $0.538_{\pm  0.013}$        &       0.587   \\
HD3360  &       B2IV    &       3.67*   &       0.090   &       $-0.618_{\pm  0.038}$        &       \cite{2MASS}    &       0.278   &       $0.350_{\pm  0.004}$        &       0.299   \\
HD33959 &       A9V     &       5.00*   &       0.006   &       $0.610_{\pm  0.026}$        &       \cite{2MASS}    &       0.476   &       $0.515_{\pm  0.004}$        &       0.472   \\
HD35468 &       B2V     &       1.64*   &       0.003   &       $-0.686_{\pm  0.038}$        &       \cite{Ducati}   &       0.267   &       $0.785_{\pm  0.007}$        &       1.162   \\
HD58142 &       A0.5V   &       4.61    &       0.006   &       $0.033_{\pm  0.018}$        &       \cite{2MASS}    &       0.410   &       $0.419_{\pm  0.008}$        &       0.303   \\
HD886   &       B2IV    &       2.83*   &       0.034   &       $-0.756_{\pm  0.032}$        &       \cite{Ducati}   &       0.277   &       $0.435_{\pm  0.004}$        &       0.478   \\
HD89021 &       A1IV    &       3.43    &       0.006   &       $0.045_{\pm  0.094}$        &       \cite{Ducati}   &       0.453   &       $0.757_{\pm  0.007}$        &       0.389   \\
HD97633 &       A2IV    &       3.32    &       0.003   &       $0.017_{\pm  0.094}$        &       \cite{2MASS}    &       0.431   &       $0.769_{\pm  0.010}$        &       1.443   \\
\hline
\end{tabular}
\\
\tablefoot{From left to right: the name of the star, its spectral type taken from the SIMBAD Astronomical Database, the $V$ magnitude \citep{Kharchenko}, the visual interstellar extinction \citep{Stilism1, Stilism2}, the $(V-K)_0$ color corrected from the extinction, the reference used for the infrared $K$ photometry, the limb-darkening coefficient, the limb-darkened angular diameter, and the reduced chi-squared.
\tablefoottext{*}{Variable stars.}}
\end{table*}

We observed the sample of 18 early-type stars %the 18-stars sample
from 23 February 2019 to 16 December 2020. The calibrated oifits files are available on OIdB\footnote{\url{http://oidb.jmmc.fr/index.html}}. 
The data were processed using the standard VEGA pipeline \citep{Mourard2009} and the squared visibilities were calibrated using reference stars selected with the SearchCal tool\footnote{The tool is available at \url{https://www.jmmc.fr/english/tools/proposal-preparation/search-cal/}} \citep{Bonneau06}. The list of the calibrators is included in Table \ref{Cals_data}. The uniform-disk angular diameter in the $R$ band $\theta_{\mathrm{UD}}$[R] is taken from the JMMC Stellar Diameters Catalogue version 2 \citep[JSDC2]{2017yCat.2346....0B}, but we consider the uncertainties from the JDSC version 1 \citep{2010yCat.2300....0L}, which is more conservative. The observing log is given in Table~\ref{observing_table}. The systematic uncertainties stem from the uncertainties on the calibrator diameters, given in Table~\ref{Cals_data}, and they are negligible with respect to the statistical ones.

We used the JMMC LitPro\footnote{The tool is available at \url{https://www.jmmc.fr/english/tools/data-analysis/litpro/}} tool \citep{LITpro} to fit a model of a linear limb-darkened disk on the calibrated squared visibilities. This model has the following two parameters: the limb-darkened angular diameter and the limb-darkening coefficient, $u_R$. Considering the precision of the VEGA measurements and the range of spatial frequencies that have been covered, we cannot adjust the coefficient of the limb darkening. The $u_{R}$ coefficient for each star is fixed and taken from the \citet{Claret} catalog. We searched for the effective temperature $T_{\mathrm{eff}}$, the gravity $\log g$, and the metallicity $Z$ of the star and we took the closest value of each parameter available in the catalog in order to determine $u_{R}$. Claret's grids have a step of 250$\,$K in temperature, thus the largest error we can make on the temperature is 125$\,$K without any interpolation. This error on the temperature leads to an average error of 0.05\% on the angular diameter, which is well below our typical errors. 

This parameter is fixed in the fitting process of the LITpro tool. The coefficient $u_{R}$ is given in Table \ref{VEGA_data} together with the derived limb-darkened angular diameter. The corresponding visibility curves are presented in Fig. \ref{vis_fig}. The precision on the angular diameter that we obtain ranges from 0.78\% to 5.10\%, with a median value of 1.8\%.

\begin{figure}
   \centering
   \includegraphics[width=\hsize]{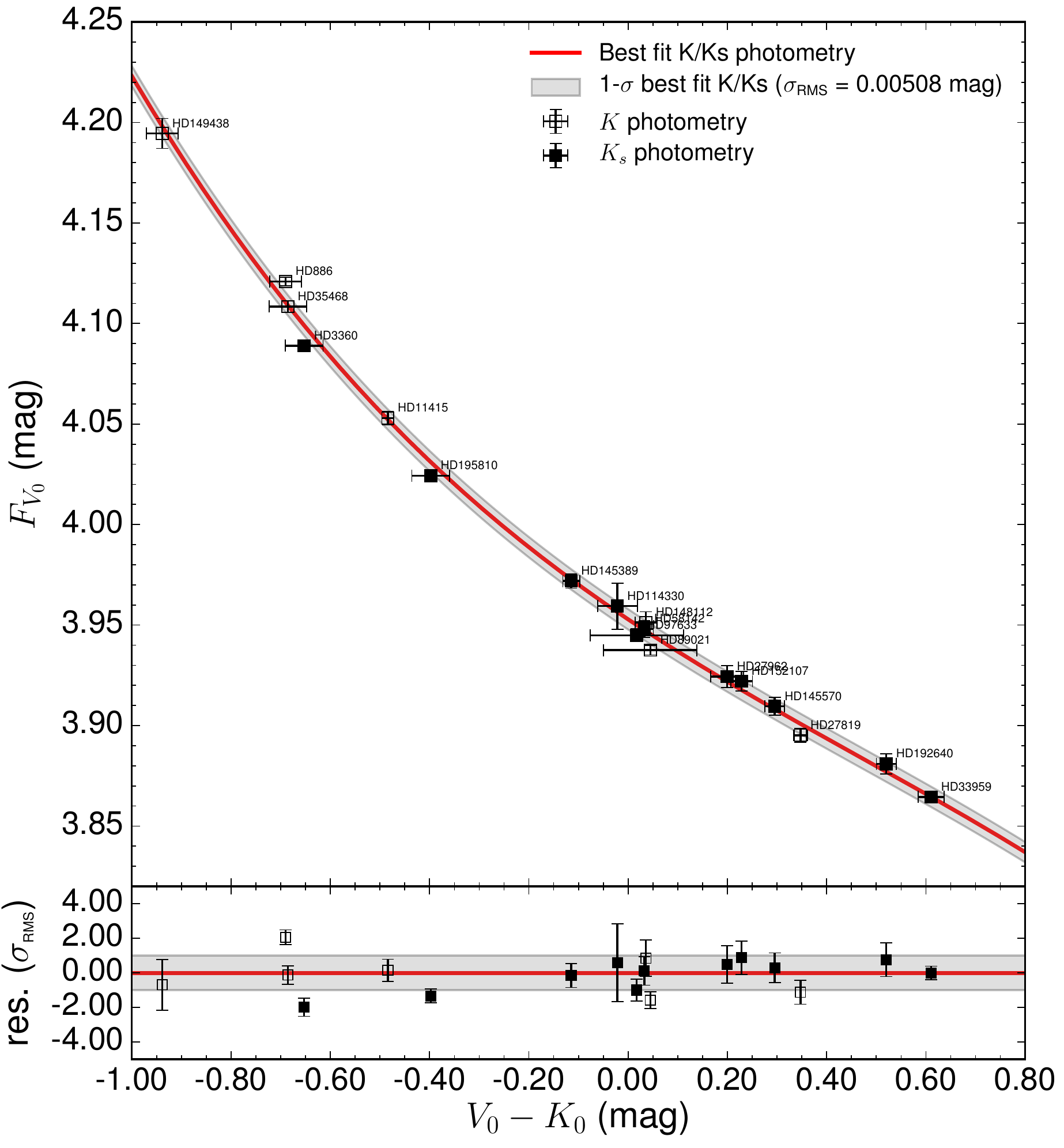}
   \caption{$F_{V_{0}}$ derived in this work as a function of $(V-K)_0$. Empty squares are data with a $K$ Johnson photometry, while filled squares are those with a $K_s$ photometry. The third-order SBCR for early-type dwarfs is shown by the red solid line. The gray area denotes the RMS of the relation. The bottom panel presents the residual in units of the RMS of the relation. See Sect. \ref{SBCR_early_type} for a description of the fitting strategy.}\label{SBC_fig}
\end{figure}

\begin{figure}
   \centering
   \includegraphics[width=\hsize]{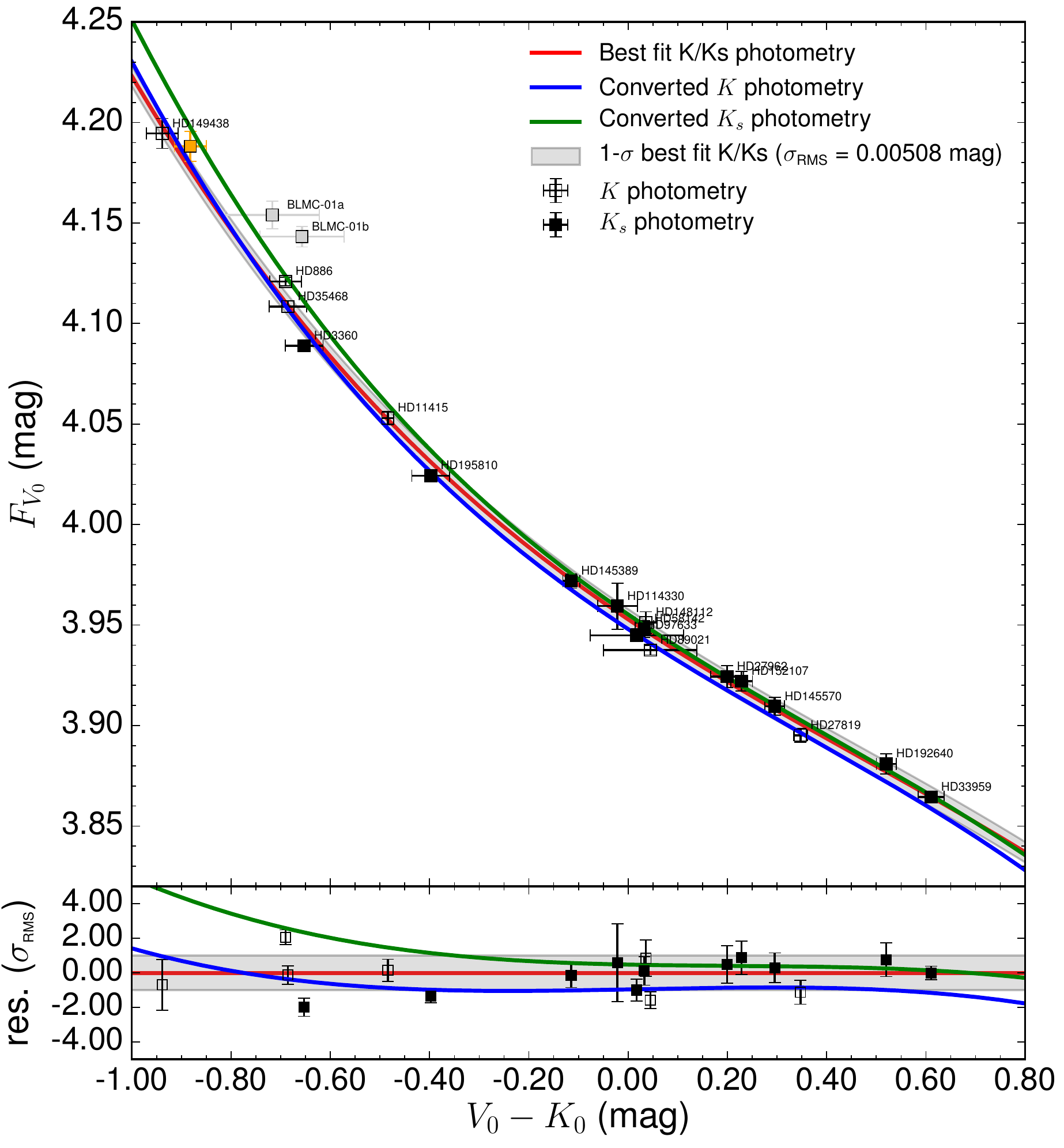}
   \caption{Comparison between our SBCR with mixed Johnson-2MASS photometries (red solid line), SBCR with converted 2MASS $K_s$ photometry (green solid line), and Johnson-$K$ photometry (blue solid line) for early-type stars. Light-gray dots are eclipsing binaries from \cite{2019ApJ...886..111T}. The orange dot shows the data for HD149438 using the Hipparcos parallax. See Sect. \ref{photometries_conv} for detailed information on the converted SBCRs.}
   \label{conv_fig}
\end{figure}

\section{Calibration of the SBCR for early-type stars}\label{SBCR_early_type}

The surface brightness $S_{\lambda}$ of a star is correlated to its limb-darkened angular diameter $\theta_{\mathrm{LD}}$ and its apparent magnitude corrected from the extinction $m_{\lambda_{0}}$ by the following formula \citep{SBCR_Wesselink}

\begin{equation}\label{4.2207}
S_{\lambda} = m_{\lambda_{0}} + 5 \log \theta_{\mathrm{LD}}.
\end{equation}

\citet{SBCR_Wesselink} used this definition to highlight the correlation between the surface brightness and the color of the star by the relation below:

\begin{equation}\label{general_SBCR}
S_{\lambda_{1}} = \sum_{n=0}^{N} C_n (m_{\lambda_{1}}-m_{\lambda_{2}})_0^n,
\end{equation}

which defines the so-called SBCR. Later, \cite{1976MNRAS.174..489B} developed another definition of the SBCR, which we denote as $F_{\lambda}$. To be consistent with the strategy of Paper I, we consider the \cite{1976MNRAS.174..489B} definition in the rest of our study, namely

\begin{equation}\label{precise_SBCR}
F_{\lambda_{1}} = 4.2196 - 0.1 \times \left\{ \sum_{n=0}^{N} C_n (m_{\lambda_{1}}-m_{\lambda_{2}})_0^n\right\}.
\end{equation}

The 18-star sample covers a range of $V-K$ color from -1 to +0.6$\,$mag. We computed the surface brightness of the 18 stars following Eq.~\ref{precise_SBCR}. Our fitting strategy uses the orthogonal distance regression (ODR), which considers both $F_V$ and $V-K$ errors. A more detailed description can be found in Appendix A of Paper I. The final SBCR for early-type stars combining  Johnson and 2MASS photometries ($K$/$K_s$ SBCR hereafter) is then shown in Fig. \ref{SBC_fig}. The coefficients of our SBCR are shown in the first row of Table \ref{SBCRs_coeff_early}, together with their uncertainties. We discuss the possibility of converting all the photometric measurements into the same system in Sect. \ref{photometries_conv}. We restricted our SBCR to the third order.

\begin{table*}
%\begin{longtable}{cccccccccccccc}
\caption{Parameters of $K/K_s$ and converted SBCRs for early-type stars (see Sect. \ref{photometries_conv} for a description of the conversions).}
\centering
\begin{tabular}{cccccccc}
\hline
\hline
                &       $N_c$ /$N$      &       $C_0$   &       $C_1$   &       $C_2$   &       $C_3$   &       $\sigma_{\mathrm{RMS}}$         &       Expected $\frac{\sigma{\theta_{\mathrm{LD}}}}{\theta_{\mathrm{LD}}}$    \\
                &               &               &               &               &               &       [mag]   &       [\%]    \\
\hline                                                                                                                          
        $K/K_s$ &               &       $2.6675_{\pm 0.0149}$   &       $1.6556_{\pm 0.0571}$        &       $-0.6084_{\pm 0.0667}$  &       $0.4350_{\pm 0.1530}$   &       0.00508 &       2.34    \\
        $K$     &       11/18   &       $2.7162_{\pm 0.0196}$   &       $1.6267_{\pm 0.0773}$        &       $-0.6191_{\pm 0.1011}$  &       $0.5733_{\pm 0.2276}$   &       0.00770 &       3.55    \\
        $K_s$   &       7/18    &       $2.6434_{\pm 0.0199}$   &       $1.6845_{\pm 0.0792}$        &       $-0.6947_{\pm 0.0924}$  &       $0.5734_{\pm 0.2276}$   &       0.00770 &       3.55    \\
\hline
\end{tabular}\label{SBCRs_coeff_early}
\tablefoot{The coefficients $C_n$ follow the definition of Eq. \ref{precise_SBCR}. The $N_c/N$ column stands for the number of converted photometries as a fraction of the total number of stars in the sample. The RMS and the corresponding expected precision on the angular diameter, computed from Eq. \ref{accuracy_equation}, are shown in the last columns.}
\end{table*}

Using Eq. \ref{4.2207} with visible $V$ magnitudes, we have:

\begin{equation}\label{equation_theta}
    \theta_{\mathrm{LD}} = 10^{8.4392-0.2 V_0 - 2 F_{V _0}}.
\end{equation}

Applying the partial derivative method on Eq. \ref{equation_theta} gives

\begin{equation}\label{accuracy_equation}
\frac{\sigma_{\theta_{\mathrm{LD}_\mathrm{rms}}}}{\theta_{\mathrm{LD}}} = 2 \ln(10) \sigma_{F_{V_{0}}}.
\end{equation}

The average RMS of the relation is found to be $\sigma_{F_{V_{0}}} = 0.00508 \,$mag.
This corresponds to a relative precision on the angular diameter of 2.3\%, according to Eq. \ref{accuracy_equation}. However, the lack of measurements in the blue part (i.e., $V-K<-0.2\,$mag) has to be taken into account. We split the $V-K$ validity domain into two ranges, namely $-0.8\,$mag $< V-K < -0.2\,$mag and $-0.2\,$mag $< V-K < 0.6\,$mag, and we computed the RMS of the SBCR on both ranges. For $V-K < -0.2\,$mag, we expect a precision of 3.1\% on the angular diameter. On the other hand, the expected precision is 1.8\% for $V-K>-0.2\,$mag.

\section{Discussion}\label{discussion}

\subsection{Impact of the variability on the SBCR}\label{variability}

As in Paper I, we aimed to quantify the impact of the variability on the SBCR. In our sample of early-type stars, ten out of 18 stars are flagged as variables. Their variability ranges from $\pm 0.01\,$mag to $\pm 0.06\,$mag according to \cite{2017ARep...61...80S}, with a median value at 0.04$\,$mag. We computed a SBCR considering an offset on the $V$ magnitude of the ten variable stars corresponding to their maximum amplitude. The resulting SBCR is consistent at a level of less than 1$\sigma$ with the current SBCR all over the $V-K$ validity domain. We conclude that a variability under 0.06$\,$mag does not have any consequence on our result. Keeping these variable stars in the sample is therefore acceptable.

\subsection{The $K$-band photometry of early-type stars}\label{photometries_conv}

The Johnson-$K$ and 2MASS-$K_s$ photometries of our sample are not equally distributed in terms of V-K colors (see Fig. \ref{SBC_fig}), which prevents one from calibrating a purely homogeneous SBCR, which is either based on $K$ or $K_s$, respectively. We therefore made the choice of combining $K$ and $K_s$ data to calibrate the SBCR. The precision we expect on the angular diameter using this relation is 2.3\%, but we do not exclude a bias due to the fact that we mixed different infrared photometric bands. However, using the conversion relations $K$ to $K_s$ or the reverse also implies some difficulties. Indeed, the typical transformation equations \citep{1988PASP..100.1134B, 2001AJ....121.2851C} are indirectly deduced from the 2MASS-CIT and 2MASS-SAAO equations of \cite{1988PASP..100.1134B} and \cite{2001AJ....121.2851C}. Moreover, the CIT equation of \cite{1988PASP..100.1134B} is based on the observation of only 21 stars, and the bluer spectral type of the sample is B7. Regarding our sample, this corresponds to a V-K color of $-0.3\,$mag.  Nevertheless, the resulting converted SBCRs are shown as blue and green solid lines in Fig. \ref{conv_fig} for Johnson-$K$ and 2MASS-$K_s$ photometries respectively. The coefficients are shown in Table \ref{SBCRs_coeff_early}. The converted Johnson-$K$ SBCR is consistent with the inhomogeneous SBCR at less than $1\sigma$ over all the $V-K$ validity domain. Concerning the uniform 2MASS relation, such a conversion does not influence the calibration of the SBCR at more than 1$\sigma$ for V-K > -0.4$\,$ mag. The inconsistency, however, reaches more than 4$\sigma$ for the bluest part of the relation. The expected precision on the angular diameter using a SBCR based on a uniform set of photometry is of 3.6\%. 

In conclusion, if one wants to derive the angular diameter of a star with $K$ photometry, we suggest using the $K$/$K_s$ or $K$ SBCRs in Table \ref{SBCRs_coeff_early}, with a good level of confidence, the $K$/$K_s$ relation being the most precise. If instead one has $K_s$ photometry for his or her star, using the SBCR based on $K_s$ is more consistent, but a bias due to the conversion from $K$ to $K_s$ in the calibration process of the SBCR is not excluded, as already discussed. Future investigations are needed to evaluate the consistency of such photometric conversion relations on the early-type stars' color range.

\subsection{Comparison with the literature}\label{comp_lite}

HD35468 was already observed by \cite{2014AA...570A.104C} and they obtained an angular diameter of 0.715 $\pm$ 0.005$\,$mas, whereas our measurement is 0.786 $\pm$ 0.007$\,$mas. 
\cite{2014AA...570A.104C} took the uniform-disk angular diameter of their calibrators in the JSDC \citep{2010yCat.2300....0L}, while we used the second version of the catalog \citep{2017yCat.2346....0B}. We processed the measurements of \cite{2014AA...570A.104C} using the JSDC2 angular diameters measurements for the three calibrators of \cite{2014AA...570A.104C}, and with the same strategy described in this work. We found an angular diameter of 0.807 $\pm$ 0.026$\,$mas, which is consistent with our value at a level of 1$\sigma$. This clearly shows some bias in the initial data reduction of \cite{2014AA...570A.104C} due to differences between JSDC and JSDC2 angular diameters.

In addition to HD35468, two other stars in Table \ref{VEGA_data} have been observed previously at CHARA. For HD97633, \cite{2013MNRAS.434.1321M} measured an angular diameter of 0.740 $\pm$ 0.024$\,$mas using PAVO, which is consistent with our result. The PAVO measurements of both \cite{2013MNRAS.434.1321M} and \cite{2019ApJ...873...91G}  led to a smaller value for the angular diameter for HD3360 (0.311 $\pm$ 0.010$\,$mas and 0.280 $\pm$ 0.018$\,$mas, respectively). The diameter measurements in both papers are based on only two observations, so they could be more susceptible to systematic errors. Also, \cite{2014MNRAS.439.2060C} and \cite{2018MNRAS.477.4403W} found increased discrepancies as the angular diameters approached the resolution limits, which is the case here with such a value for the angular diameter for HD3360.

\begin{figure}
   \centering
   \includegraphics[width=\hsize]{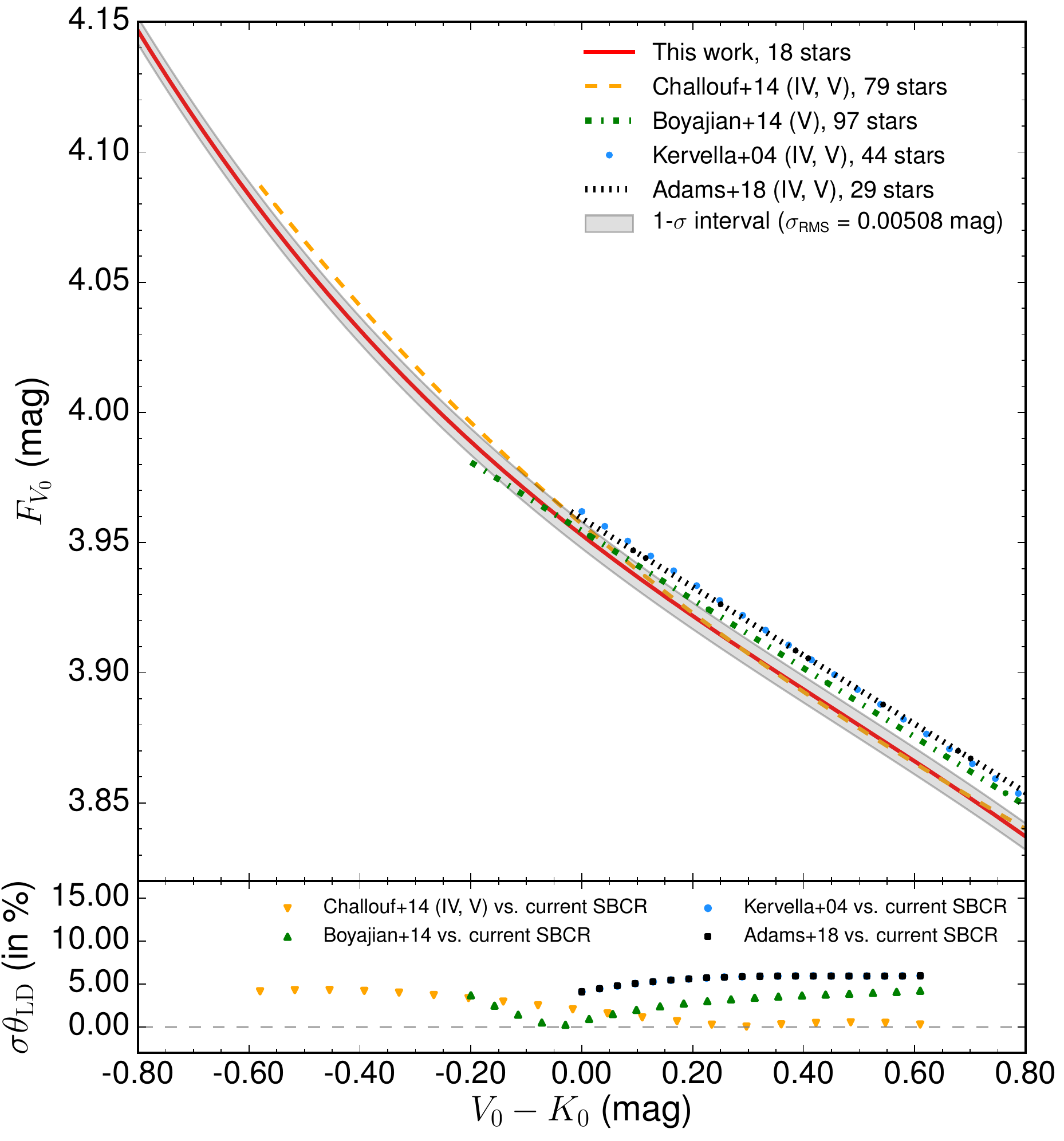}
   \caption{Top panel: Comparison between our SBCR for early-type stars and relations of \cite{2014AA...570A.104C}, \cite{Boyajian14}, \cite{Kervella04b}, and \cite{Adams2018}. Bottom panel: Differences in angular diameter estimates between our SBCR and the other relations. See Sect. \ref{comp_lite} for more details on the comparison.}
   \label{comp_SBCR}
\end{figure}

We also compared our extinction values with those of \cite{2014AA...570A.104C}. We computed the extinction of our 18 stars by combining the $Q$-method from \cite{1953ApJ...117..313J} and the intrinsic colors method by \cite{2014AcA....64..261W}. We found consistent results with \textit{Stilism}, except for one star, namely HD149438. This star has a smaller Gaia parallax value than Hipparcos, but the  values of the extinction are consistent ($A_v = 0.21\,$mag for Gaia versus $A_V = 0.15\,$mag for Hipparcos). The uncertainty of the Gaia parallax is larger than the one of Hipparcos. We suggest this star could be oversaturated in Gaia photometry broadbands. In Fig. \ref{conv_fig}, we included  the data for HD149438 using the Hipparcos parallax (orange dot). The choice of the parallax does not affect the SBCR at more than 1$\sigma$. For consistency, we decided to keep the Gaia parallax value for HD149438 in this work.

In Fig. \ref{comp_SBCR}, we included a comparison between our SBCR for early-type stars and the relations of \cite{Boyajian14}, \cite{Kervella04b}, \cite{2014AA...570A.104C}, and \cite{Adams2018}. The bottom panel shows the normalized difference we expect in terms of the angular diameter between our SBCR and these relations on the $-0.6 < V-K <$ 0.6$\,$mag color range. The first two references have been largely used so far, and their $V-K$ color domain of validity cover a large part of the early-type range. Conversely to our SBCR, these relations are fully linear. We expect a difference on the angular diameter of more than 5\% for $V-K > 0\,$mag using the relations of \cite{Kervella04b} and \cite{Adams2018}. We find an  agreement of less than 4\% on the angular diameter with the linear relation of \cite{Boyajian14} for $V-K>-0.2\,$mag. This comparison demonstrates that a linear SBCR for early-type stars is no longer valid for $V-K < -0.2\,$mag. Comparing our new relation with the dwarfs relation of \cite{2014AA...570A.104C} leads to a good agreement of less than 5\% on -0.6 $< V-K < 0.6\,$mag. The agreement is even more evident for $V-K > 0\,$mag, with a difference of at most 2\% on the angular diameter estimate.

Among the eight stars measured by \cite{2014AA...570A.104C}, six of them do not fulfill the stellar characteristics
criteria, which can also explain this gap. The inconsistency we see for $V-K < -0.2\,$mag can also be explained by the consideration of photometric uncertainties. Indeed, \cite{2014AA...570A.104C} have assumed respective errors of 0.015$\,$mag and 0.03$\,$mag on the $V$ and $K$ photometries of their sample, while, via this work, we see that the infrared $K$ photometry plays a major role in the calibration of a SBCR. The uncertainty on the $K$ photometry often exceeds 0.03$\,$mag in our sample, considering an arbitrary error could therefore induce an underestimation of the photometric uncertainty and finally a bias in the calibration of the relation.

\cite{2019ApJ...886..111T, 2020ApJ...890..137T} analyzed two early-type eclipsing binaries in the LMC with the aim to obtain precise and accurate stellar parameters of each early-type stars, which were then used to derive the surface brightness. They compare the measurements of their first binary system BLMC-01 of class IV or V,  with the all-classes relation of \cite{2014AA...570A.104C} and they find good agreement. However, their measurements are inconsistent with our inhomogeneous $K/K_s$ SBCR (see light-gray dots on Fig. \ref{conv_fig}) at more than 5.5$\sigma$. The disagreement is lower but still significant when considering the SBCR converted into $K_s$ system. Such inconsistencies require more investigation.  

\section{Updating late-type stars SBCRs and linking to early-type stars 
}\label{improvement_SBCRs}

\begin{figure*}
   \centering
   \includegraphics[width=0.5\hsize]{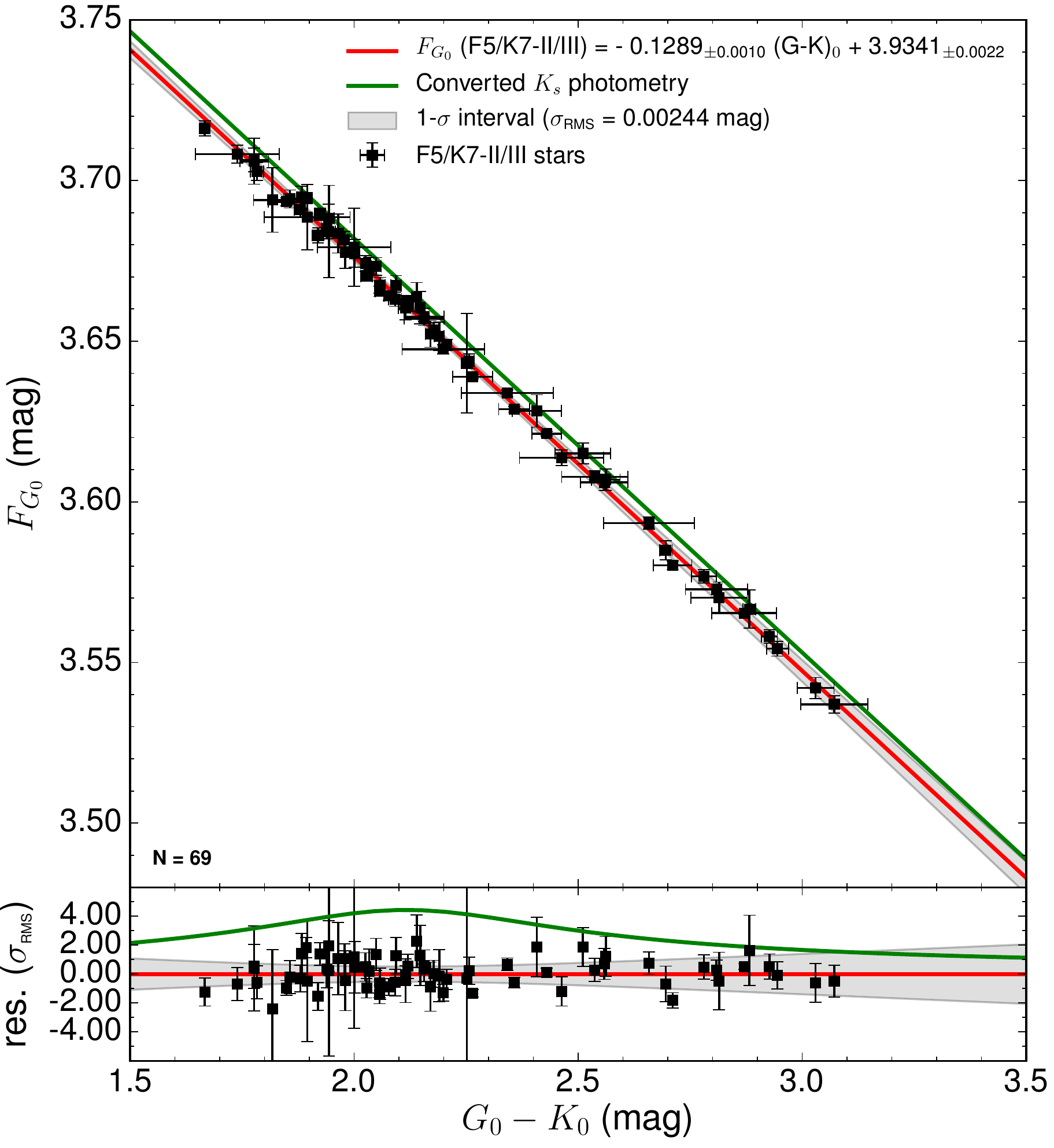}\includegraphics[width=0.50\hsize]{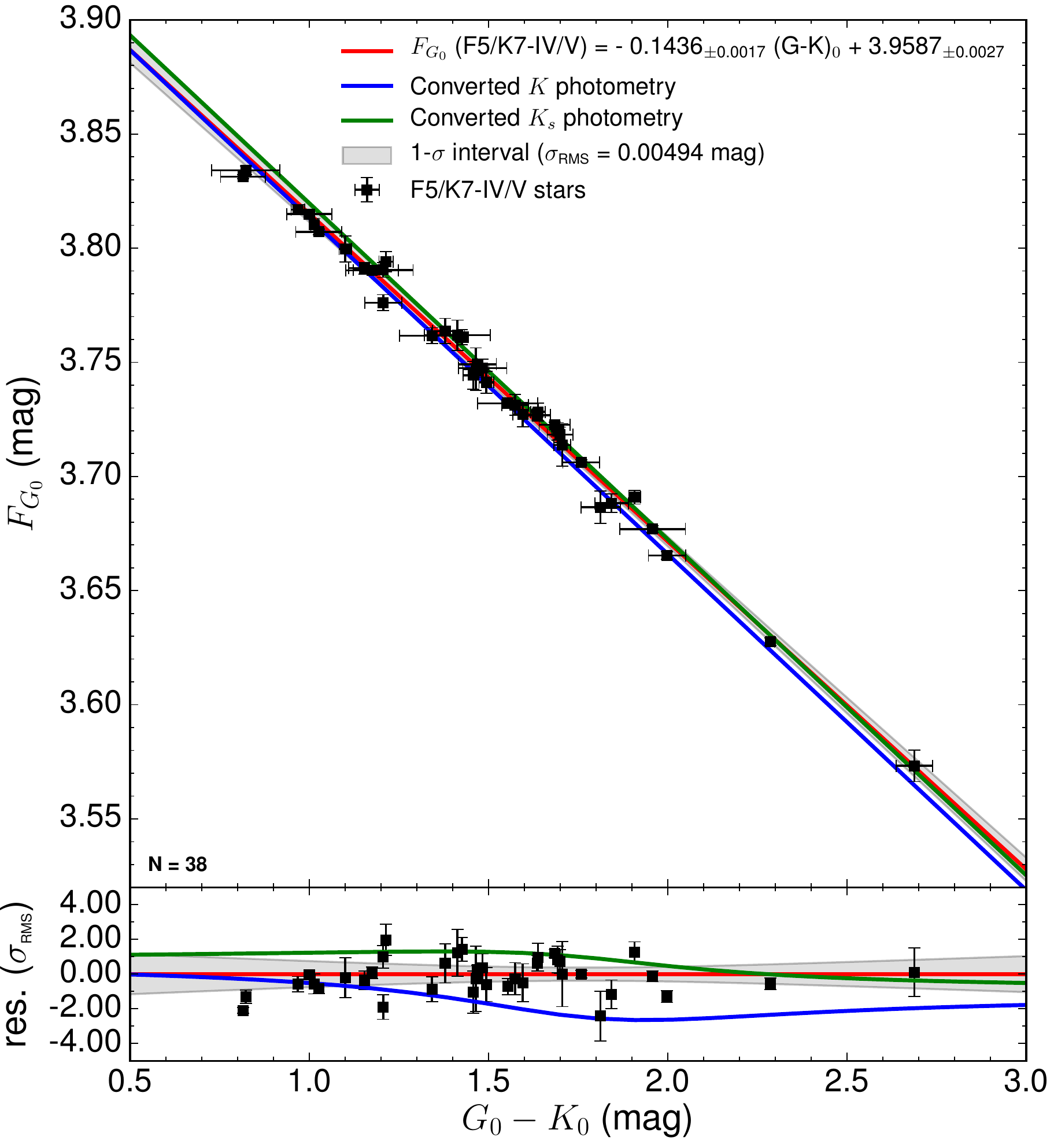}
   \includegraphics[width=0.50\hsize]{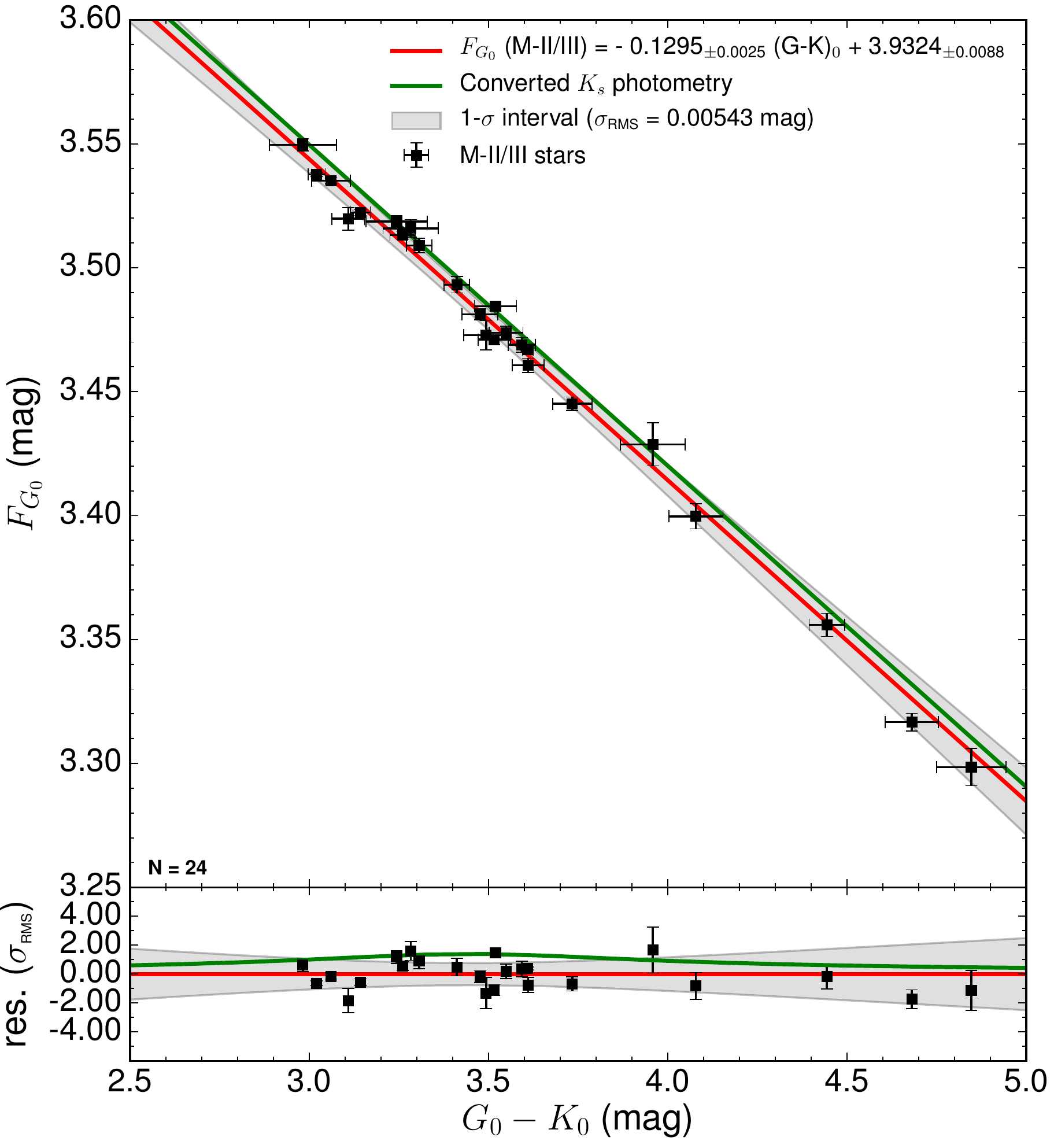}\includegraphics[width=0.50\hsize]{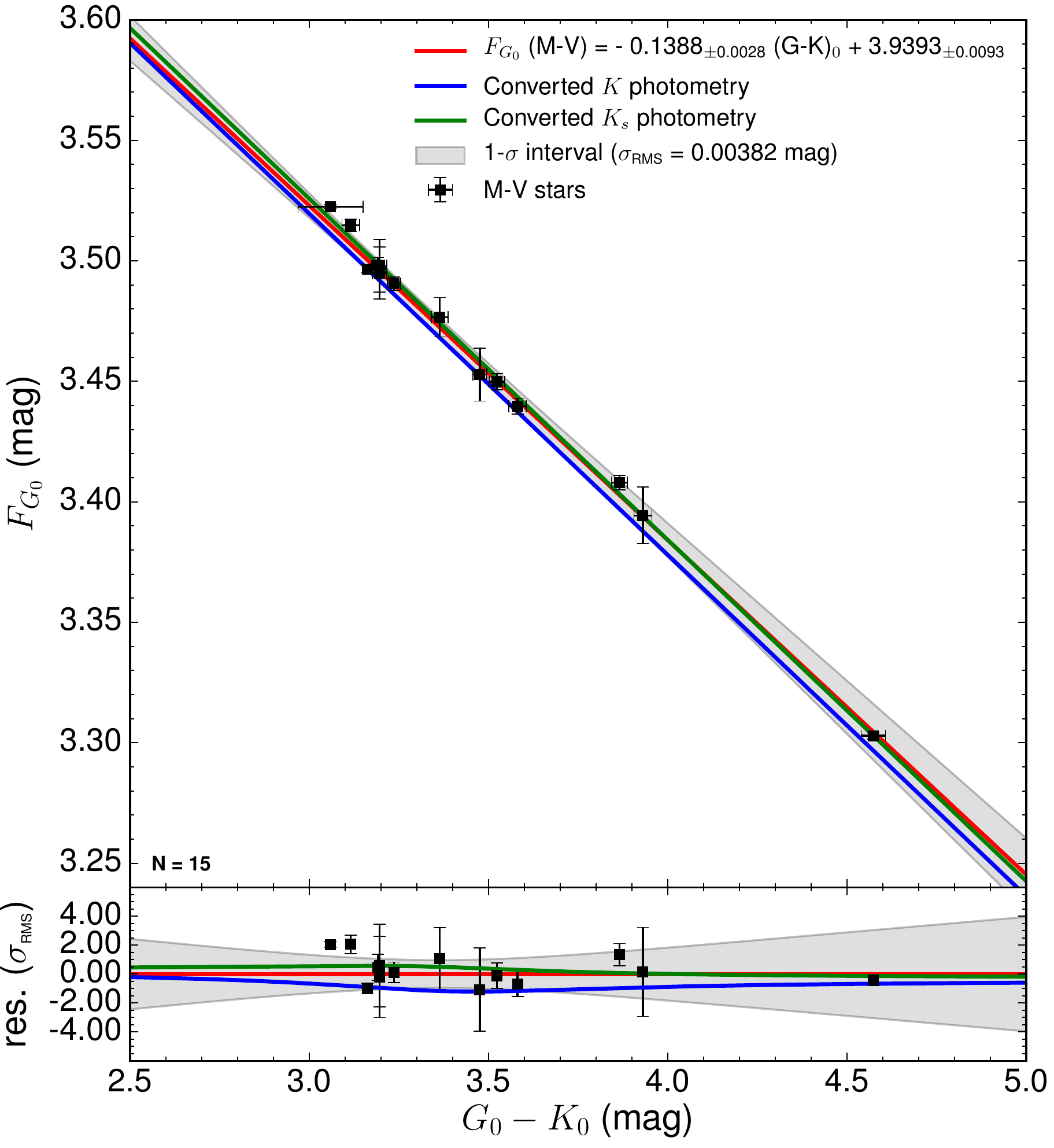}
    \caption{Converted SBCRs for late-type stars using Gaia photometry, plotted above the data (black squares). Relations with converted Johnson $K$-photometries are shown as a blue solid line, while 2MASS $K_s$ ones are in a green solid line. See Sect. \ref{section_gaia} for more information.}
         \label{homogeneous_SBCRs_Gaia}
\end{figure*}

\begin{figure*}
   \centering
   \includegraphics[width=0.5\hsize]{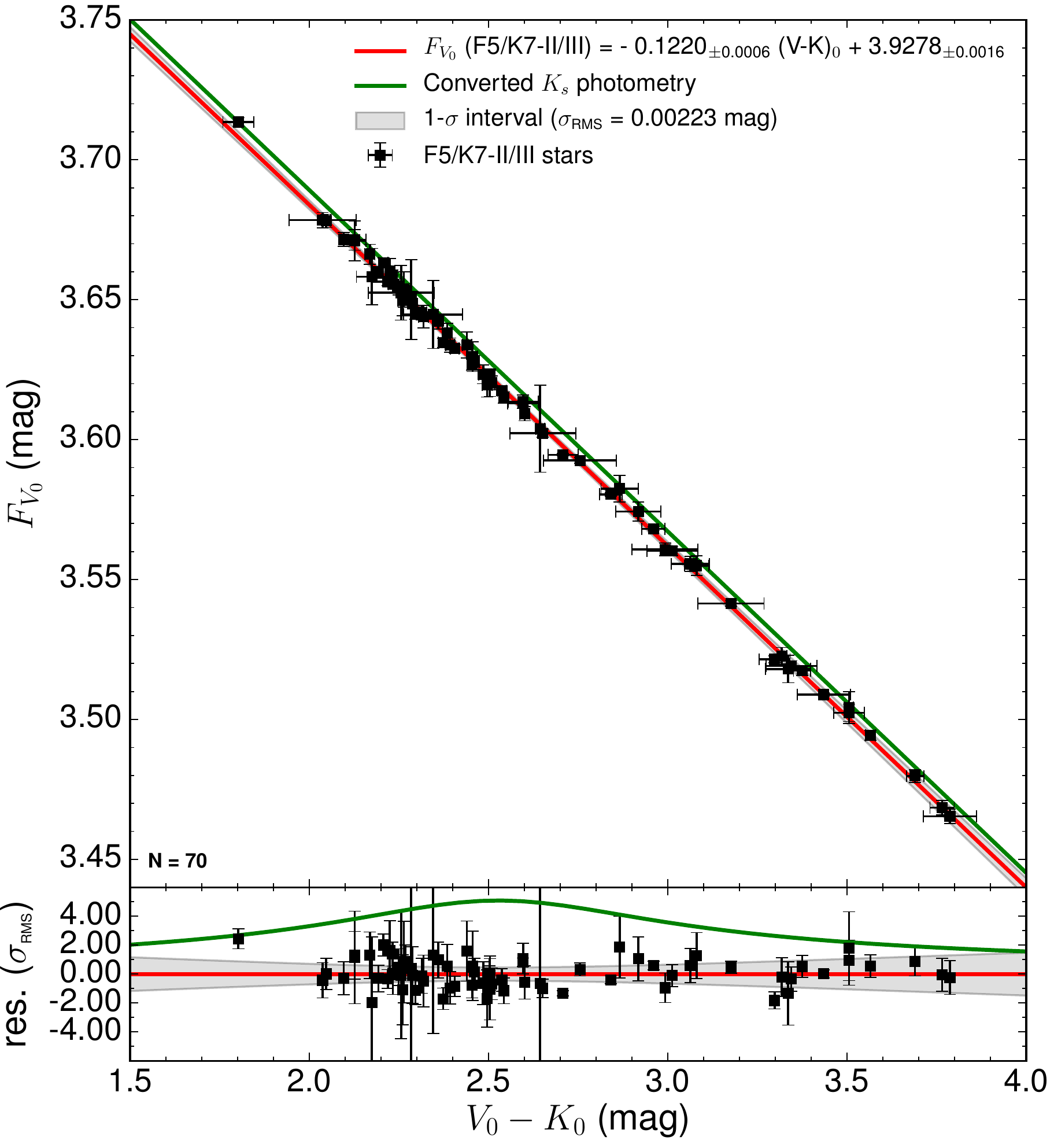}\includegraphics[width=0.50\hsize]{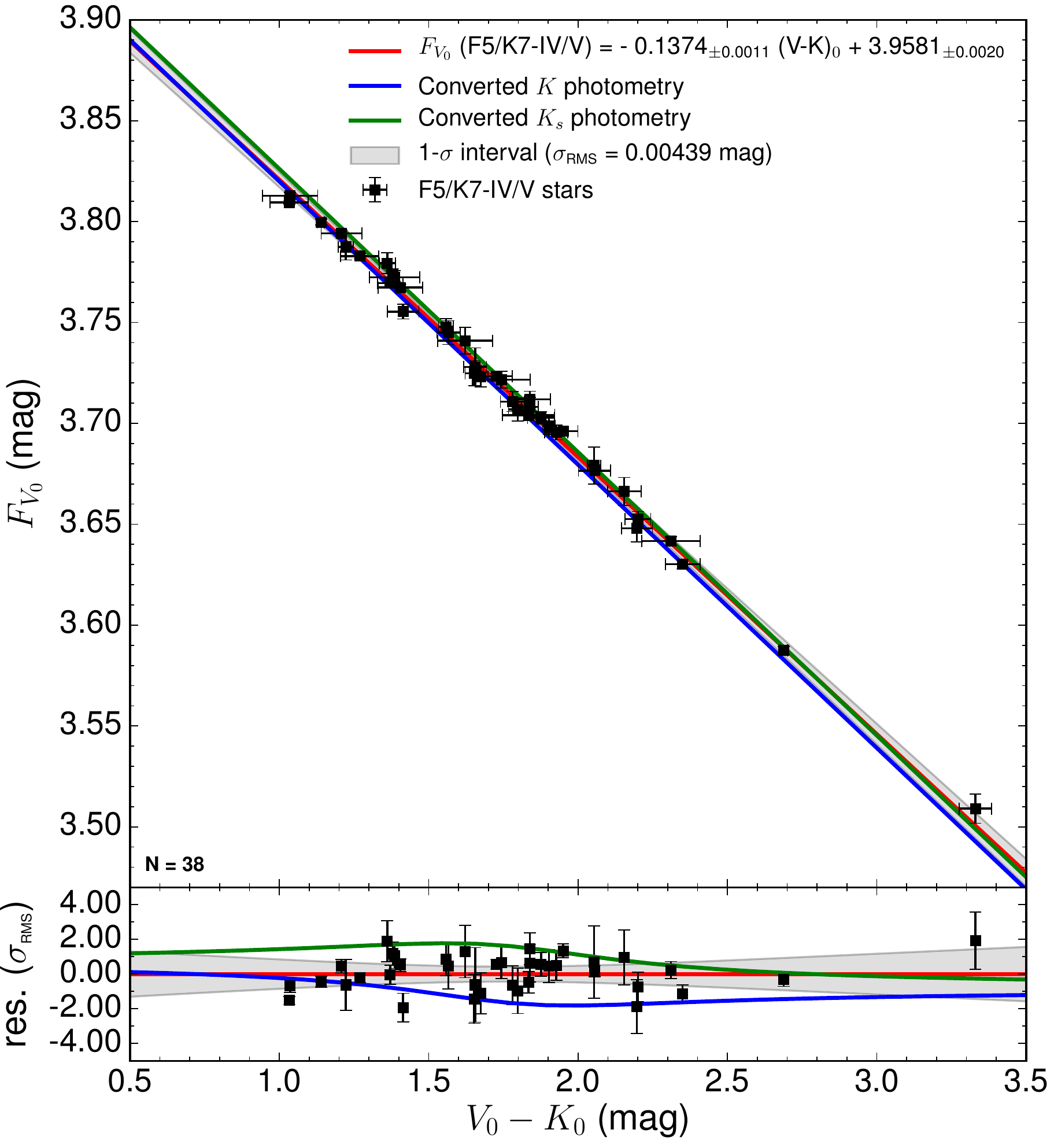}
   \includegraphics[width=0.50\hsize]{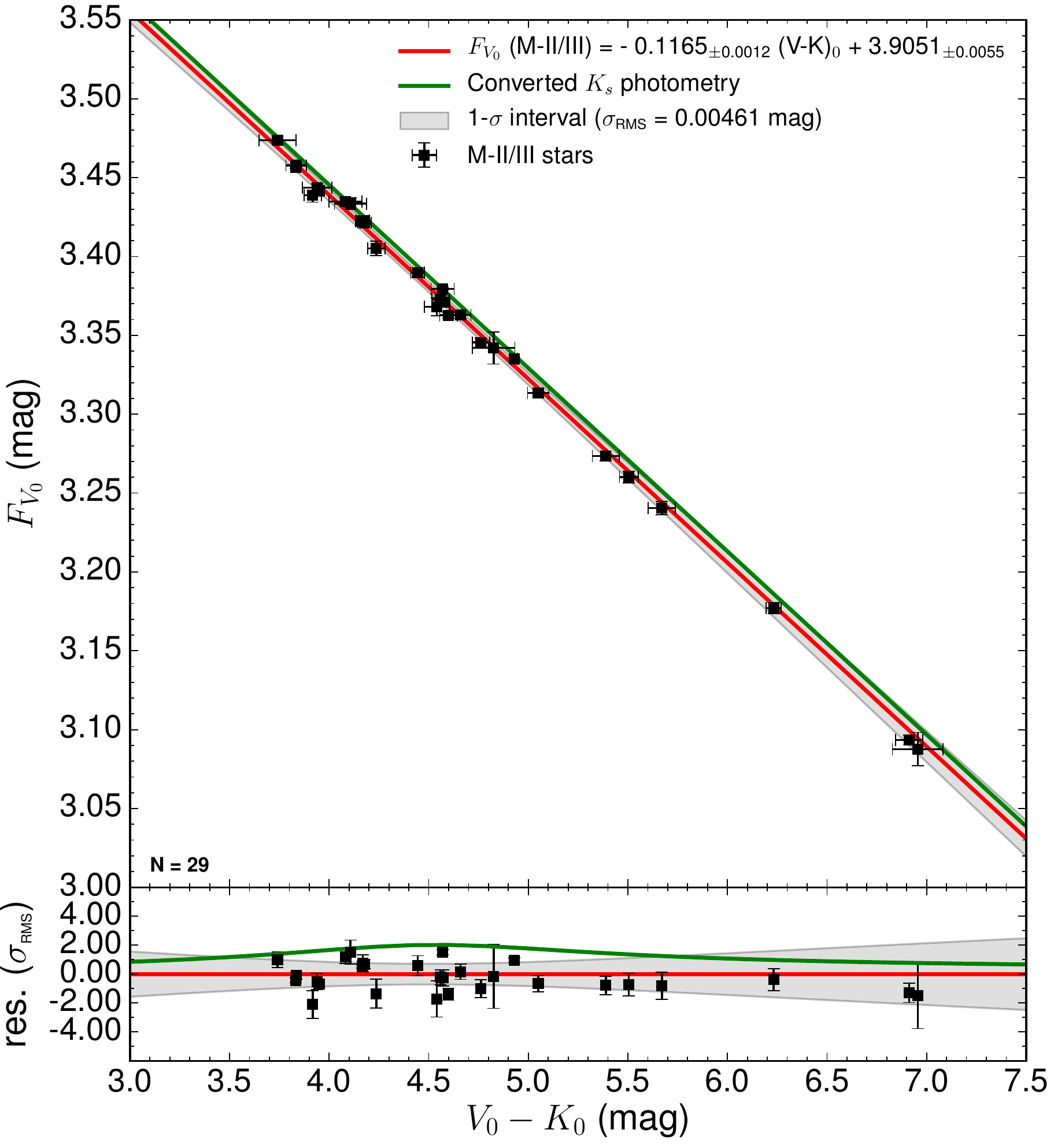}\includegraphics[width=0.50\hsize]{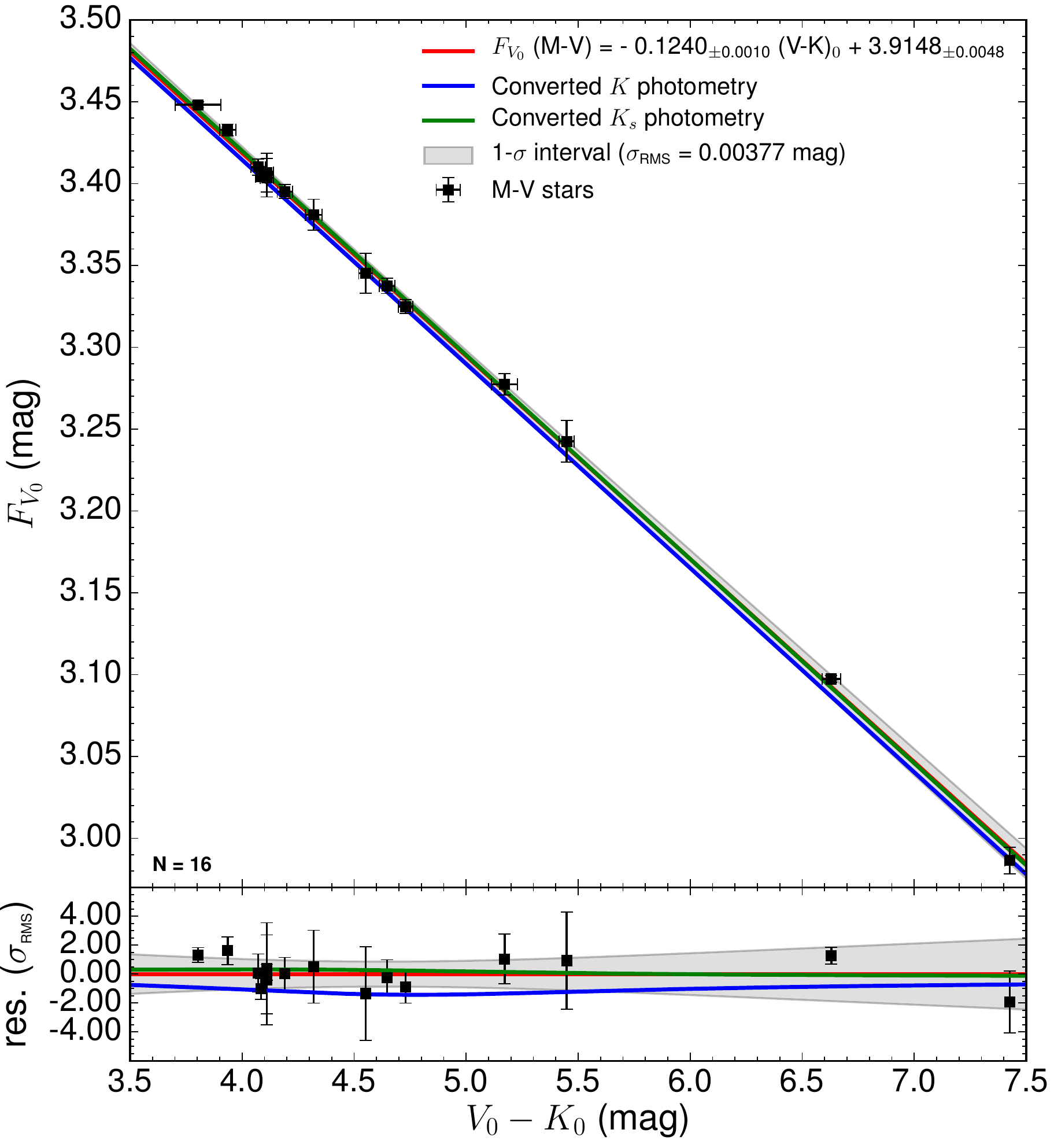}
      \caption{SBCRs for late-type stars using a converted photometric set for each sample, plotted above the data (black squares). Relations with converted Johnson $K$-photometries are shown as a blue solid line, while 2MASS $K_s$ ones are in a green solid line. See Sect. \ref{section_gaia} for more information.}
         \label{homogeneous_SBCRs}
\end{figure*}

\subsection{Update on late-type SBCRs}\label{section_gaia}

In Paper I, we implemented SBCRs using Gaia photometry. We noticed an error in the extinction calculation for the $G$-band. Indeed, Eq. 11 from Paper 1 should have been rewritten in the following way \citep{Danielski}:

\begin{multline}
%\begin{split}
A_G =  A_V \times (a_1 + a_2 (G-K)_0 + a_3 (G-K)_0^2 + a_4 (G-K)_0^3 \\
 + a_5 A_V + a_6 A_V^2+ a_7 (G-K)_0 A_V),
%\end{split}
\end{multline}

with $a_1 = 0.935556283, a_2 = -0.090722012, a_3 = 0.014422056, a_4 = -0.002659072, a_5 = -0.030029634, a_6 = 0.000607315,$ and $a_7 = 0.002713748$. The converted SBCRs using the Gaia photometry are shown in Fig. \ref{homogeneous_SBCRs_Gaia}, together with their parameters in Table \ref{homogeneous_SBCRs_table_Gaia}. The expected precision on the angular diameter ranges from 1.1\% to 2.5\%.

We also propose to improve the calibration of our SBCRs for late-type stars obtained in papier I by using transformation equations to build a uniform infrared photometry set. We used Eq. A1 from \cite{2001AJ....121.2851C}, namely $K_s = K - 0.044\,$mag, to transform our photometric sets. The resulting SBCRs are shown in Fig. \ref{homogeneous_SBCRs} for each sample (i.e., F5/K7-II/III, F5/K7-IV/V, M-II/III, and M-V stars). Their parameters are included in Table \ref{homogeneous_SBCRs_table}. The angular diameter precision ranges from 1.0\% to 2.7\%.

\begin{table*}
%\begin{longtable}{cccccccccccccc}
\caption{Parameters of the new converted SBCRs for late-type stars calibrated with Gaia photometry.}
\centering
\begin{tabular}{ccccccc}
\hline
\hline
	&		&	$N_c$ /$N$	&	SBCR $F_{G_{0}}$ vs. $(G-K)_0$ & $(G-K)$ range	&	$\sigma_{\mathrm{RMS}}$ 	&	Expected $\frac{\sigma{\theta_{\mathrm{LD}}}}{\theta_{\mathrm{LD}}}$	\\
	&		&		&	& [mag]	&	[mag]	&	[\%]	\\
\hline											
F5/K7-II/III	&  Paper I\tablefootmark{*} &		&	$F_{G_{0}} = -0.1289_{\pm 0.0010} (G-K)_0 + 3.9341_{\pm 0.0022}$ & 	[1.6; 3.1] &	0.00244	&	1.12	\\
	&	$K$	&	0/69	&	$F_{G_{0}} = -0.1289_{\pm 0.0010} (G-K)_0 + 3.9341_{\pm 0.0022}$	& [1.6; 3.1] &	0.00244	&	1.12	\\
	&	$K_s$	& 69/69	&	$F_{G_{0}} = -0.1289_{\pm 0.0010} (G-K_s)_0 + 3.9398_{\pm 0.0022}$	& [1.7; 3.2] &	0.00244	&	1.12	\\
\hline											
F5/K7-IV/V	&	Paper I\tablefootmark{*}	&		&	$F_{G_{0}} = -0.1436_{\pm 0.0017} (G-K)_0 + 3.9587_{\pm 0.0027}$	& [0.8; 2.7] &	0.00494	&	2.27	\\
	&	$K$	&	16/38	&	$F_{G_{0}} = -0.1472_{\pm 0.0015} (G-K)_0 + 3.9604_{\pm 0.0025}$	& [0.8; 2.7] & 0.00494	&	2.27	\\
	&	$K_s$	&	22/38	&	$F_{G_{0}} = -0.1472_{\pm 0.0015} (G-K_s)_0 + 3.9669_{\pm 0.0024}$	& [0.8; 2.8] &	0.00494	&	2.27	\\
\hline											
M-II/III	& 	Paper I\tablefootmark{*}	&		&	$F_{G_{0}} = -0.1295_{\pm 0.0025} (G-K)_0 + 3.9324_{\pm 0.0088}$ & [2.9; 4.9]	&	0.00543	&	2.50	\\
	&	$K$	&	0/24	&	$F_{G_{0}} = -0.1295_{\pm 0.0025} (G-K)_0 + 3.9324_{\pm 0.0088}$	& [2.9; 4.9] &	0.00543	&	2.50	\\
	&	$K_s$	&	24/24	&	$F_{G_{0}} = -0.1295_{\pm 0.0025} (G-K_s)_0 + 3.9381_{\pm 0.0089}$	&  [3.0; 4.9] &	0.00543	&	2.50	\\
\hline											
M-IV/V	& Paper I\tablefootmark{*}&		&	$F_{G_{0}} = -0.1388_{\pm 0.0028} (G-K)_0 + 3.9393_{\pm 0.0093}$	& [3.0; 4.6] &	0.00382	&	1.76	\\
	&	$K$	&	13/15	&	$F_{G_{0}} = -0.1416_{\pm 0.0021} (G-K)_0 + 3.9444_{\pm 0.0069}$ & [3.0; 4.6]	&	0.00407	&	1.87	\\
	&	$K_s$	&	2/15	&	$F_{G_{0}} = -0.1416_{\pm 0.0021} (G-K_s)_0 + 3.9506_{\pm 0.0069}$ & [3.0; 4.6]	&	0.00407	&	1.87	\\
\hline
\end{tabular}\tablefoot{
 The corrected relations from Paper I are shown in the first row for each sample of stars. We describe the correction applied here in Sect. \ref{section_gaia}. The $N_c/N$ column stands for the number of converted photometries as a fraction of the total number of stars in the sample. The RMS and the corresponding expected precision on the angular diameter are shown in the last columns.
\tablefoottext{*}{SBCR of Paper I corrected from the extinction error (see Sect. \ref{section_gaia}).}}\label{homogeneous_SBCRs_table_Gaia}
\end{table*}

\begin{table*}
%\begin{longtable}{cccccccccccccc}
\caption{Parameters of the SBCRs for late-type stars described in Paper I, but converted into the same photometric system (see Sect. \ref{section_gaia}).}
\centering
\begin{tabular}{ccccccc}
\hline
\hline
        &               &       $N_c$ /$N$      &       SBCR $F_{V_{0}}$ vs. $(V-K)_0$   &       $(V-K)$ range & $\sigma_{\mathrm{RMS}}$         &       Expected $\frac{\sigma{\theta_{\mathrm{LD}}}}{\theta_{\mathrm{LD}}}$     \\
        &               &               &       & [mag] &       [mag]   &       [\%]    \\
\hline                                                                                  
F5/K7-II/III    &       $K$     &       0/70    &       $F_{V_{0}} = -0.1220_{\pm 0.0006} (V-K)_0 + 3.9278_{\pm 0.0016}$  &       [1.8; 3.8] & 0.00223    &       1.03    \\
        &       $K_s$   &       70/70   &       $F_{V_{0}} = -0.1220_{\pm 0.0006} (V-K_s)_0 + 3.9332_{\pm 0.0016}$        & [1.8; 3.9]    & 0.00223       &       1.03    \\
\hline                                                                                  
F5/K7-IV/V      &       $K$     &       16/38   &       $F_{V_{0}} = -0.1404_{\pm 0.0014} (V-K)_0 + 3.9603_{\pm 0.0025}$  &       [1.0; 3.4] & 0.00575    &       2.65    \\
        &       $K_s$   &       22/38   &       $F_{V_{0}} = -0.1404_{\pm 0.0014} (V-K_s)_0 + 3.9665_{\pm 0.0025}$        & [1.0; 3.4]    & 0.00575       &       2.65    \\
\hline                                                                                  
M-II/III        &       $K$     &       0/29    &       $F_{V_{0}} = -0.1165_{\pm 0.0012} (V-K)_0 + 3.9051_{\pm 0.0055}$  & [3.7; 7.0]    & 0.00461       &       2.12         \\
        &       $K_s$   &       29/29   &       $F_{V_{0}} = -0.1165_{\pm 0.0012} (V-K_s)_0 + 3.9102_{\pm 0.0055}$        &       [3.7; 7.0] & 0.00461    &       2.12         \\
\hline                                                                                  
M-IV/V  &       $K$ &   14/16   &       $F_{V_{0}} = -0.1247_{\pm 0.0009}(V-K)_0 + 3.9133_{\pm 0.0044}$  & [3.8; 7.4]    & 0.00404       &       1.86    \\
        &       $K_s$   &       2/16    &       $F_{V_{0}} = -0.1247_{\pm 0.0009}(V-K_s)_0 + 3.9188_{\pm 0.0044}$ & [3.8; 7.4]    & 0.00404       &       1.86    \\
\hline
\end{tabular}\label{homogeneous_SBCRs_table}
\tablefoot{The $N_c/N$ column stands for the number of converted photometries as a fraction of the total number of stars in the sample. The RMS and the corresponding expected precision on the angular diameter are shown in the last columns.}
\end{table*}

\subsection{Connecting late-type to early-type SBCRs}

Our F5/K7-IV/V SBCR covers a color range from $V-K = 1\,$mag to $V-K = 2.7\,$mag, while our SBCR for early-type dwarf stars is valid up to $V-K=0.6\,$mag. There is therefore a gap in the validity domain of our relations between $V-K=0.6\,$mag and $V-K = 1\,$mag. One way to solve this problem would be to merge both samples and deduce a unique SBCR. The methodology used to develop both SBCRs differs regarding the interferometric measurements we have selected to calibrate the relations. The SBCR for early-tpe stars is a homogeneous relation calibrated using measurements from a single instrument, namely VEGA. On the other hand, SBCRs for late-type stars were implemented with measurements taken from different instruments. Also, we have shown in Paper I that SBCRs depend on the spectral type of stars, where we raised strong discrepancies between M stars and F5/K7 stars. We thus made the choice not to calibrate a single SBCR for early- and late-type stars to avoid any systematics due to a mix of spectral types, and also due to a mix of uniform and inhomogeneous sets of interferometric measurements. By extending our F5/K7-IV/V SBCR as calibrated in Paper I (based on $K$ and $K_s$ photometry) until $V-K=0.6\,$mag, we find a difference of $2\sigma$ between both early-type (using $K$/$K_s$) and late-type relations. They are however consistent in the error bars at 1$\sigma$. The same result is found when comparing consistently uniform $K$ and $K_s$ SBCRs, respectively. We expect the CHARA/SPICA instrument to solve this gap by providing more F0 to F5 measurements.

\section{Conclusions and perspectives}\label{conclusion}

We carefully selected 18 early-type stars according to selection criteria in order to measure their angular diameter with the VEGA combiner at the CHARA Array interferometer. The mean precision we obtain on our angular diameter measurements is 2\%. Using our SBCR leads to an expected statistical precision of 2.3\% on the derived angular diameter, but the user should take care of photometric conversion issues discussed in Sect \ref{photometries_conv}.

This work supports the results shown by Paper I. Indeed, to reach such statistical precision, we demonstrated the necessity of implementing selection criteria in order to calibrate the SBCRs. We also showed the need of including the $V-K$ uncertainties in the fitting process. In Paper I, we showed that SBCRs for late-type stars depend on the class of stars. In this work, and as a first step, we considered only dwarfs and subgiants stars. We also improved SBCRs for late-type stars presented in Paper I by converting the photometric data into the same system, and also by correcting the calculation of the extinction in the Gaia band.

The lack of a large set of homogeneous angular diameter interferometric measurements remains a major problem in the implementation of SBCRs. With the future SPICA instrument \citep{SPICA} that will be installed at the CHARA array, we expect to derive the angular diameter of 800 stars all over the HR diagram, with a 1\% precision level. CHARA/SPICA will also study rotation, multiplicity, wind, and environment by performing images in the visible domain. That should therefore reinforce our knowledge of SBCRs for early-type stars and in particular help to better understand the impact of stellar activity.

\begin{acknowledgements}
This work is based upon observations obtained with the Georgia State University Center for High Angular Resolution Astronomy Array at Mount Wilson Observatory.  The CHARA Array is supported by the National Science Foundation under Grant No. AST-1636624 and AST-1715788.  Institutional support has been provided from the GSU College of Arts and Sciences and the GSU Office of the Vice President for Research and Economic Development.
This work made use of the JMMC Measured stellar Diameters Catalog \citep{Duvert}. This research made use of the SIMBAD and VIZIER\footnote{Available at \url{http://cdsweb.u- strasbg.fr/}} databases at CDS, Strasbourg (France) and the electronic bibliography maintained by the NASA/ADS system. This work has made use of data from the European Space Agency (ESA) mission Gaia (\url{https://www.cosmos.esa.int/gaia}). This research has made use of the Jean-Marie Mariotti Center \texttt{OIFits Explorer} service\footnote{Available at \url{http://www.jmmc.fr/oifitsexplorer}}. This research also made use of Astropy, a community-developed core Python package for Astronomy \citep{astropy2018}. MT acknowledges financial support from the Polish National Science Center grant PRELUDIUM 2016/21/N/ST9/03310. The research leading to these results has received funding from the European Research Council (ERC) under the European Union’s Horizon 2020 research and innovation program (grant agreement No 695099) and from the National Science Center, Poland grant BEETHOVEN UMO-2018/31/G/ST9/03050. We acknowledge support from the DIR/WK/2018/09 grant of the Polish Ministry of Science and Higher Education.
\end{acknowledgements}


\bibliographystyle{aa}
\bibliography{salsi.bib}

\begin{appendix}

\section{VEGA visibility curves}

\begin{figure*}
   \centering
   \includegraphics[width=0.5\hsize]{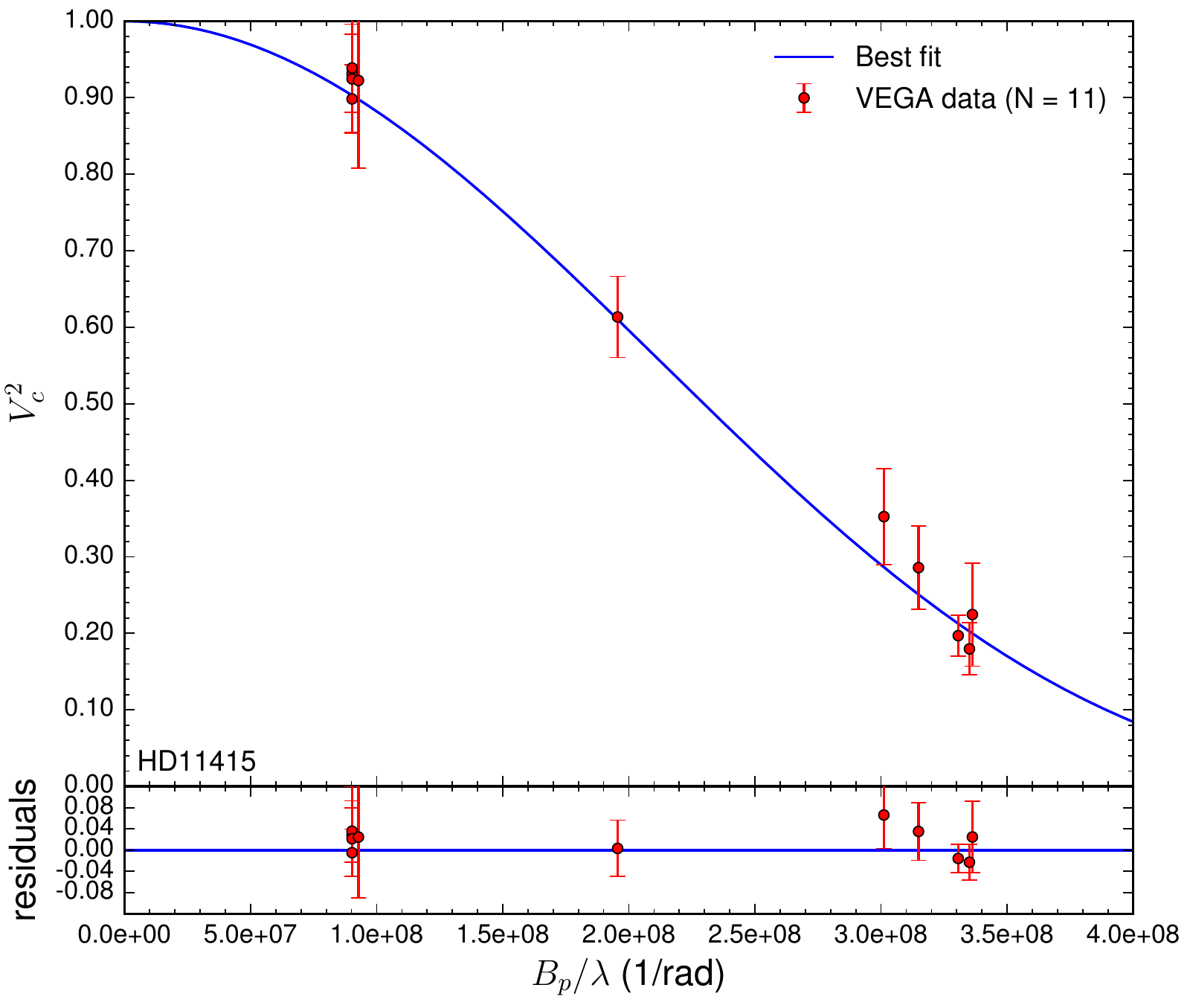}\includegraphics[width=0.5\hsize]{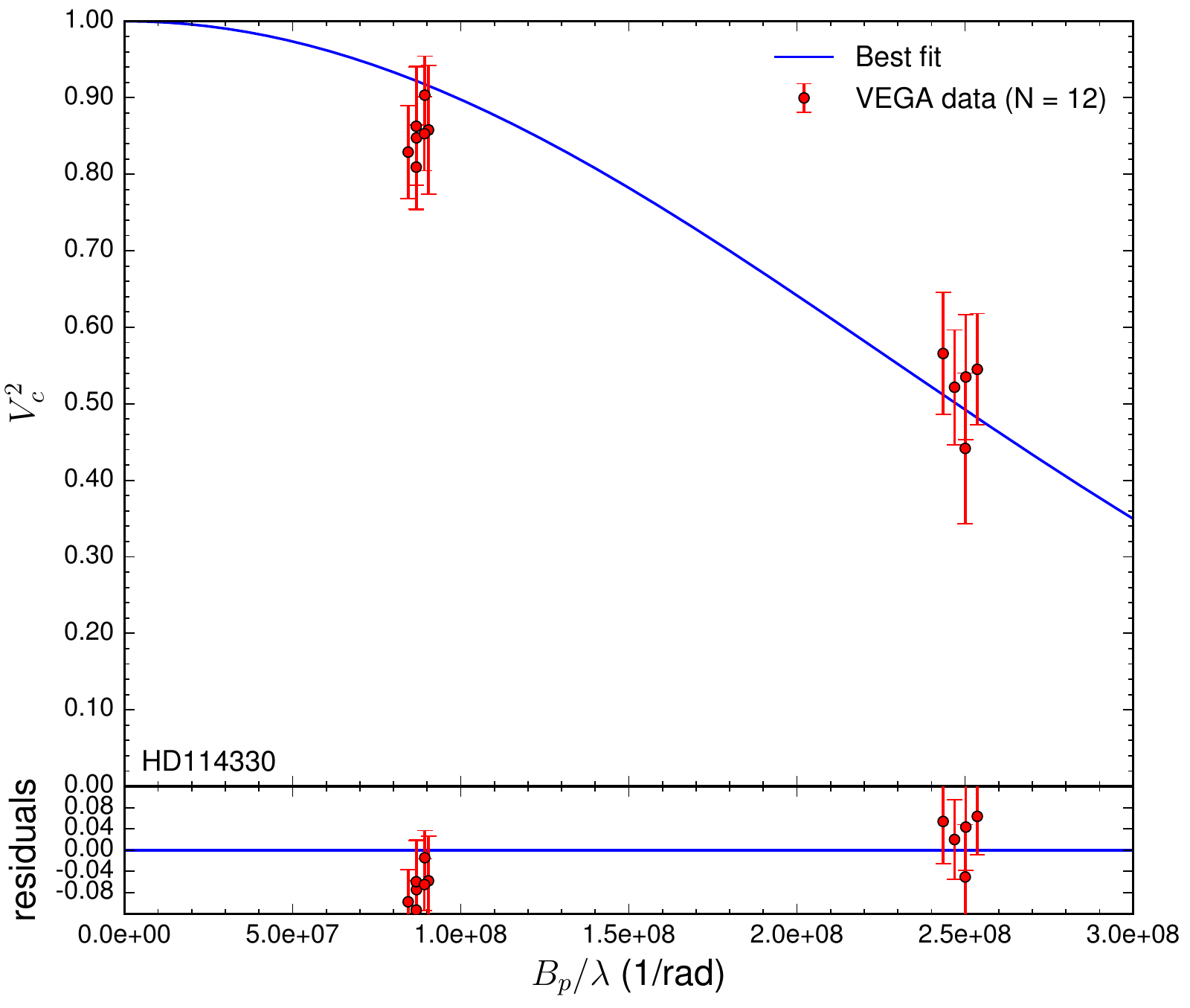}
   \includegraphics[width=0.5\hsize]{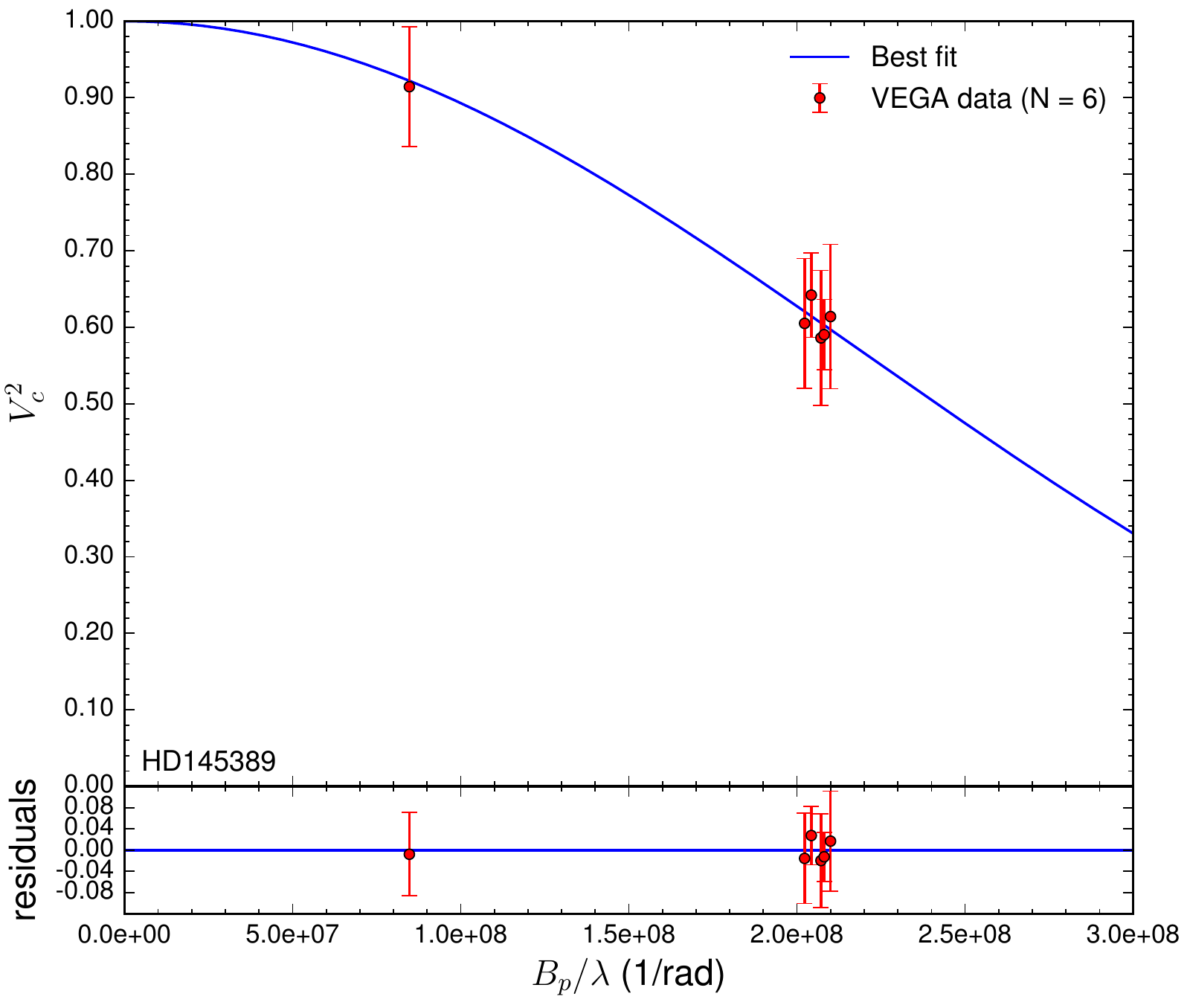}\includegraphics[width=0.5\hsize]{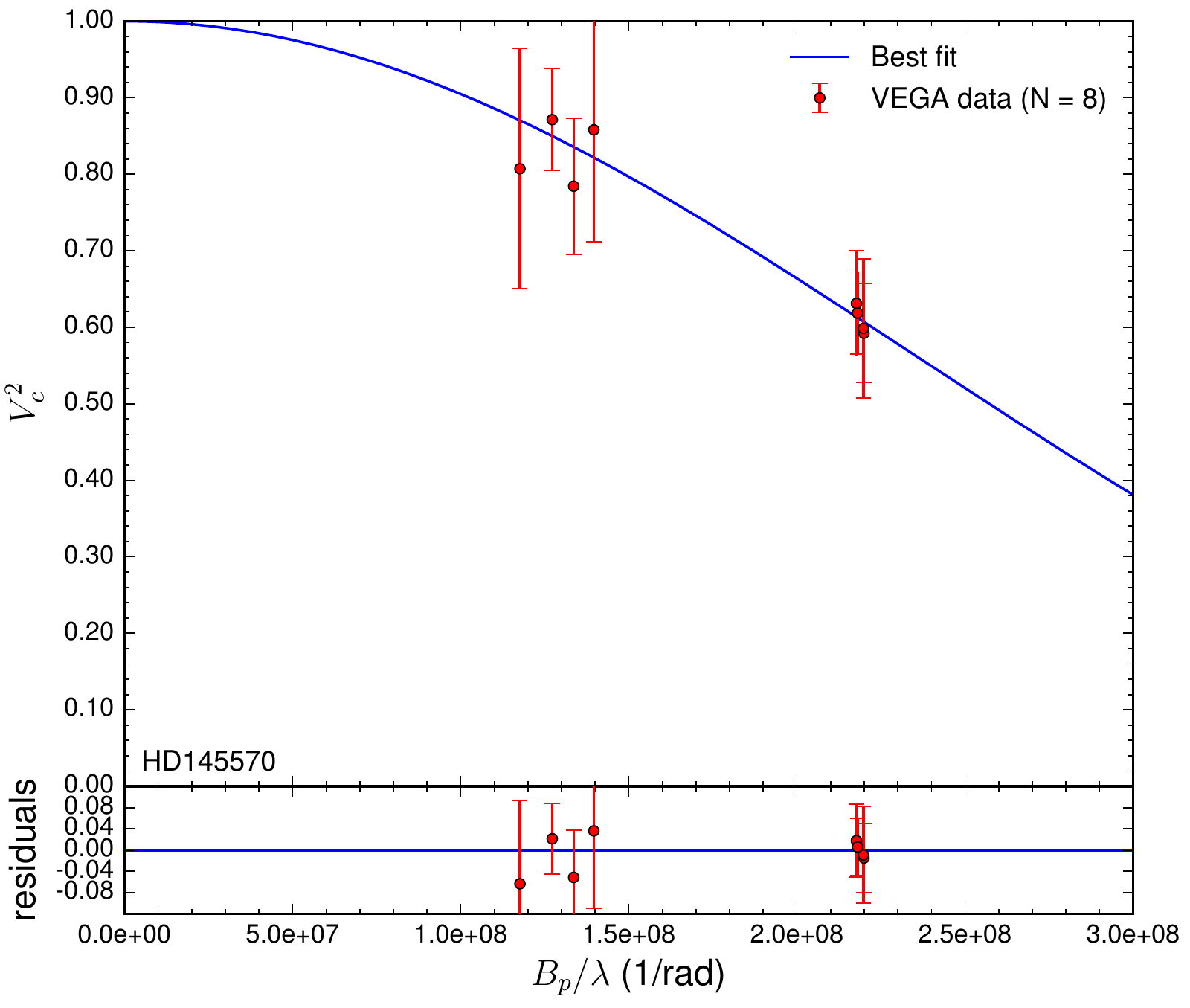}
   \includegraphics[width=0.5\hsize]{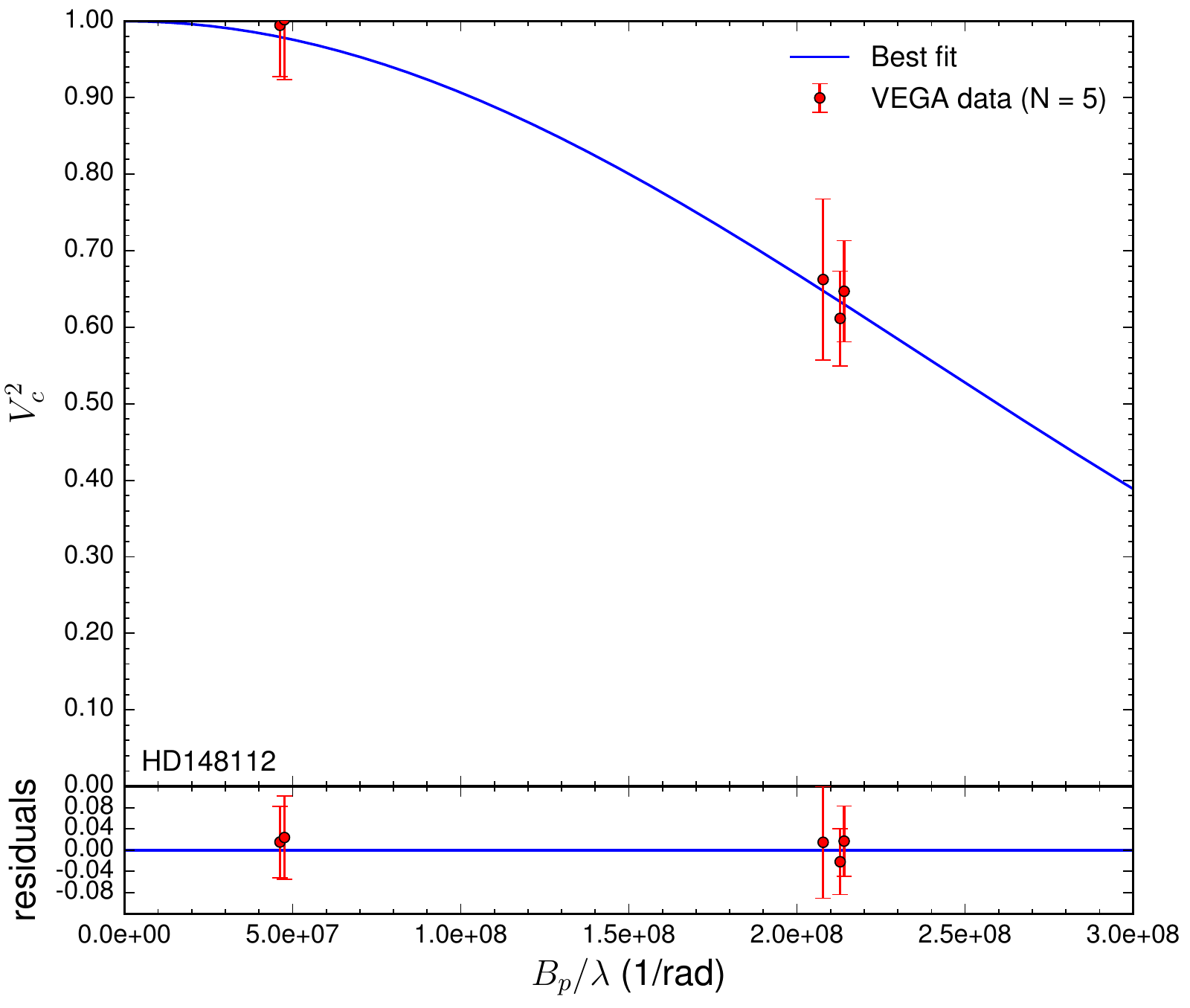}\includegraphics[width=0.5\hsize]{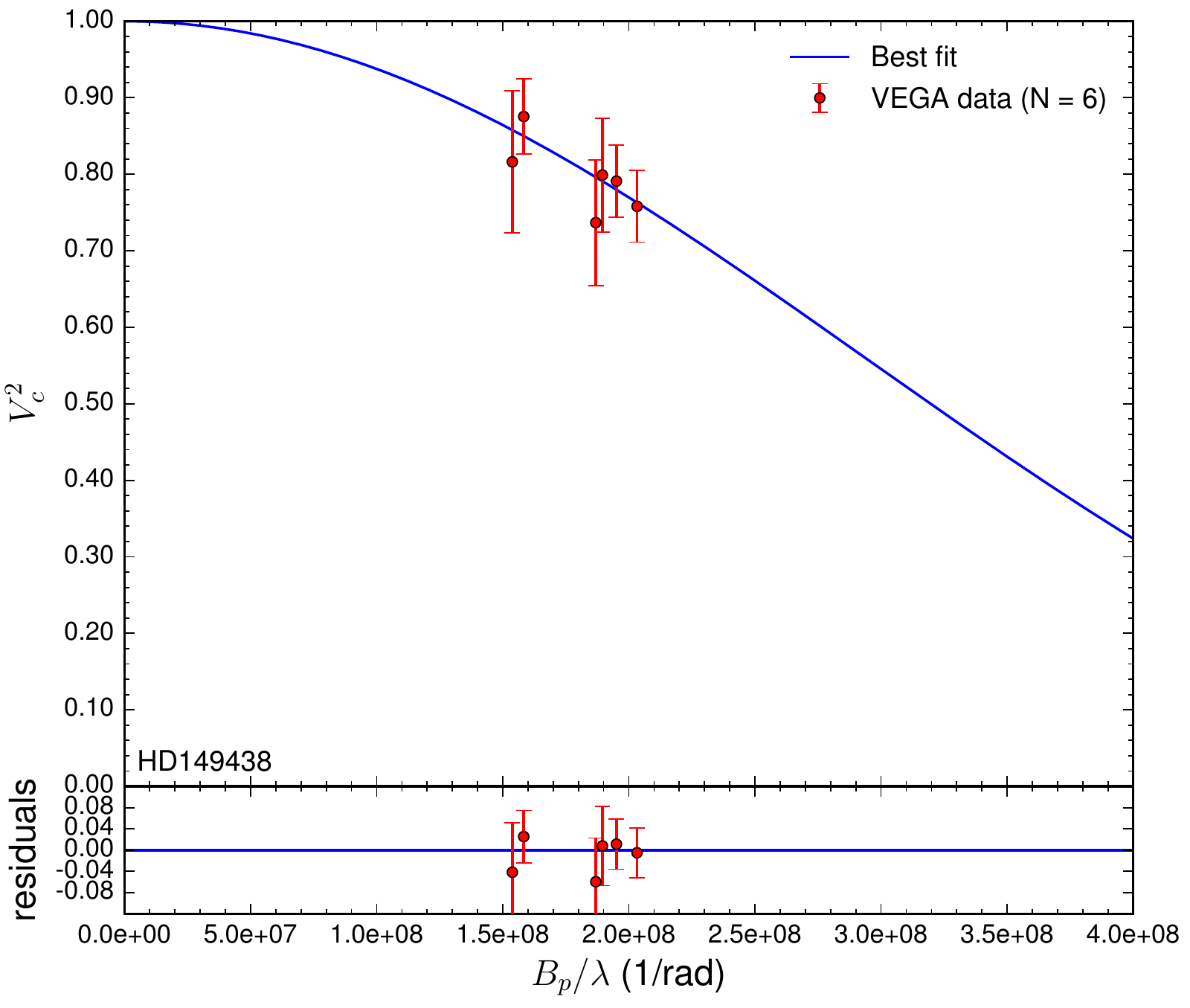}
\end{figure*}

\begin{figure*}  
   \includegraphics[width=0.5\hsize]{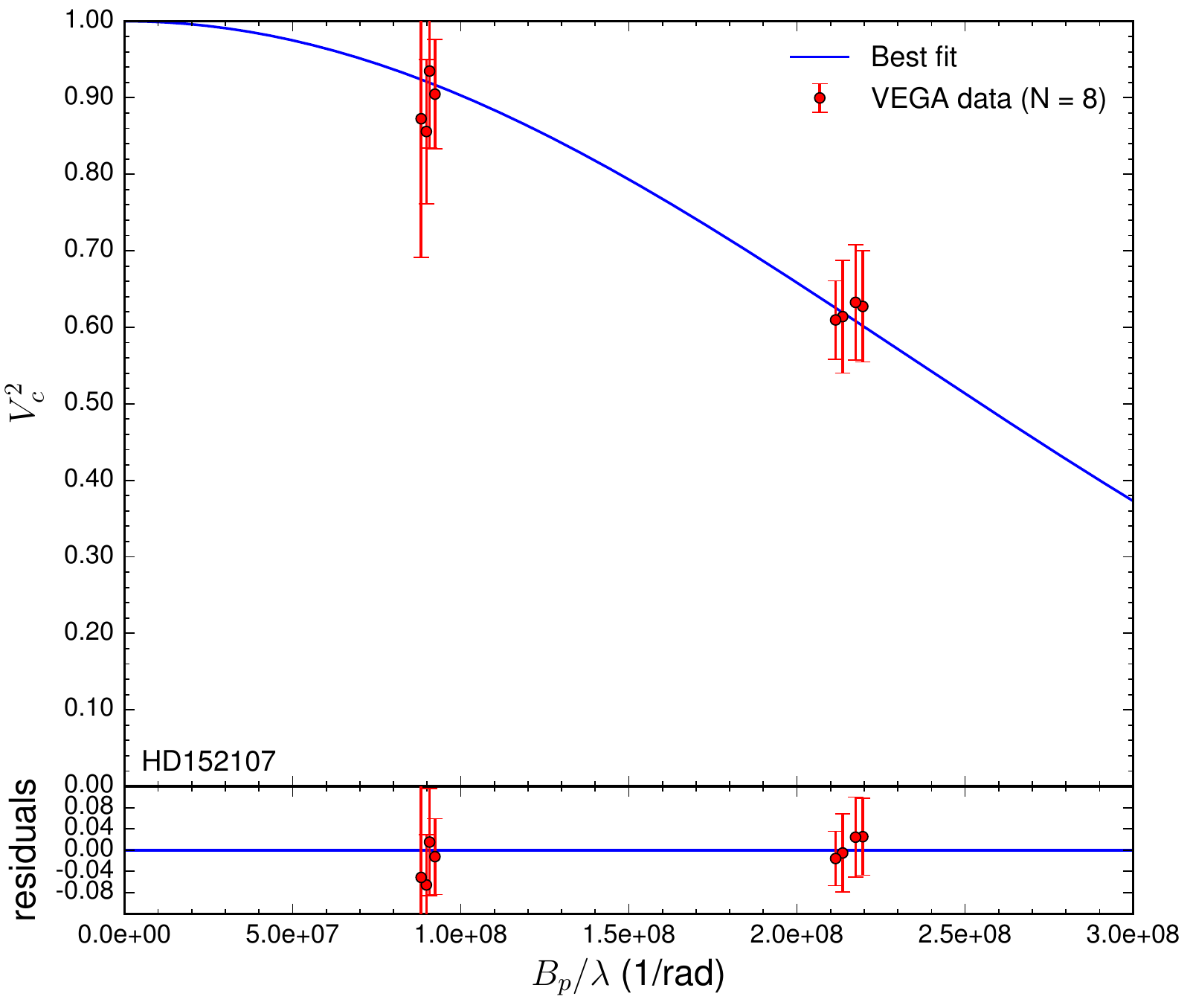}\includegraphics[width=0.5\hsize]{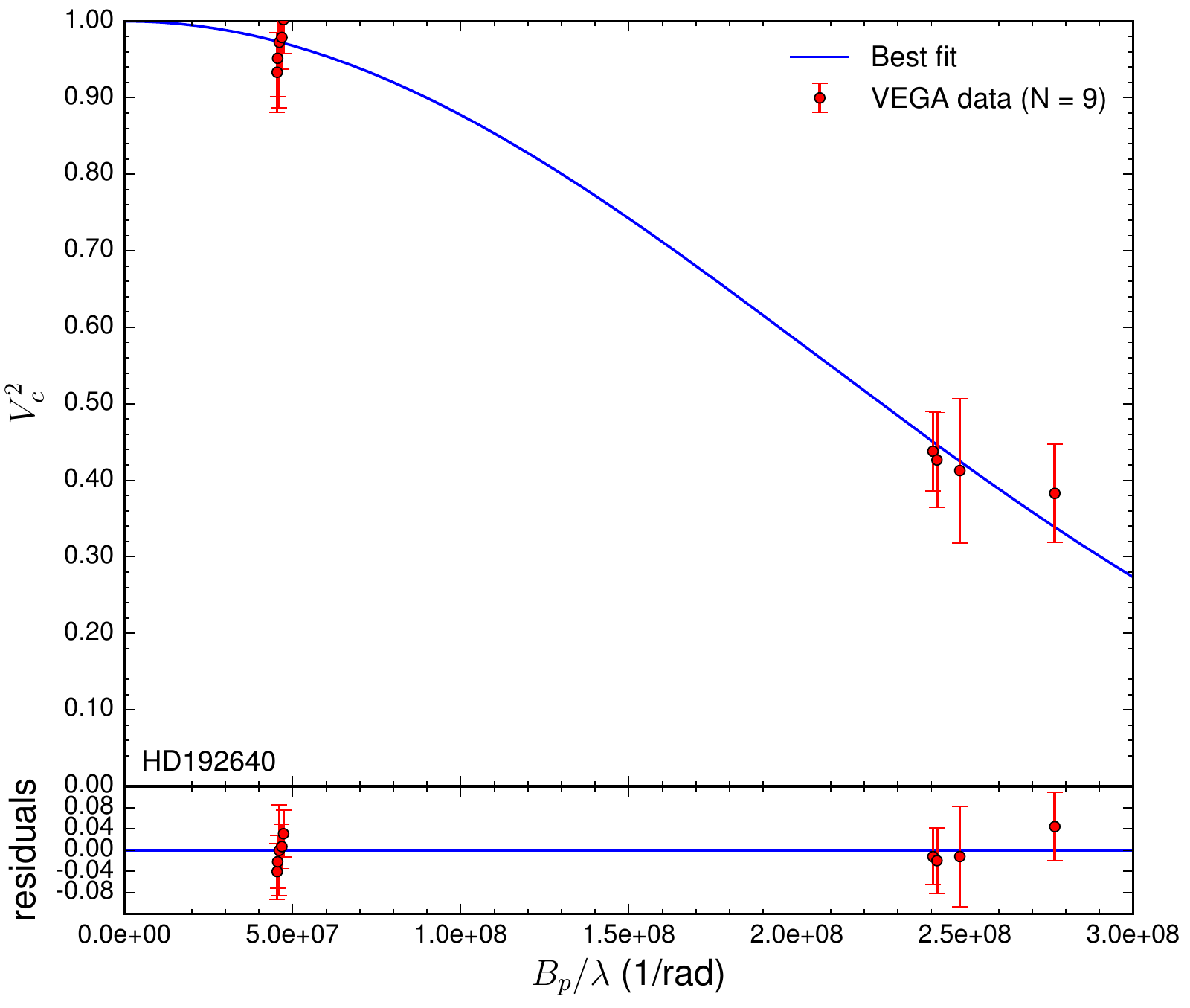}
   \includegraphics[width=0.5\hsize]{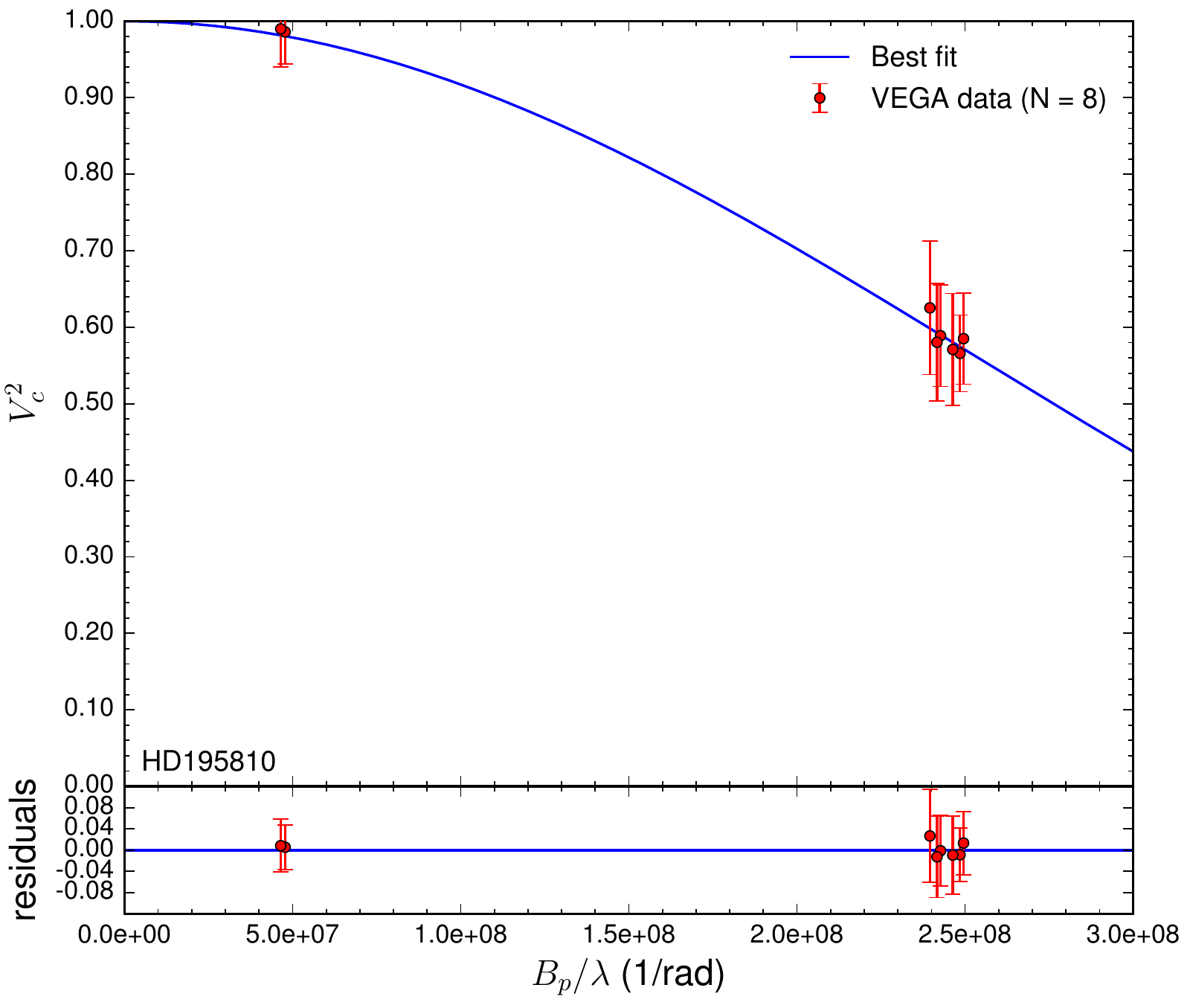}\includegraphics[width=0.5\hsize]{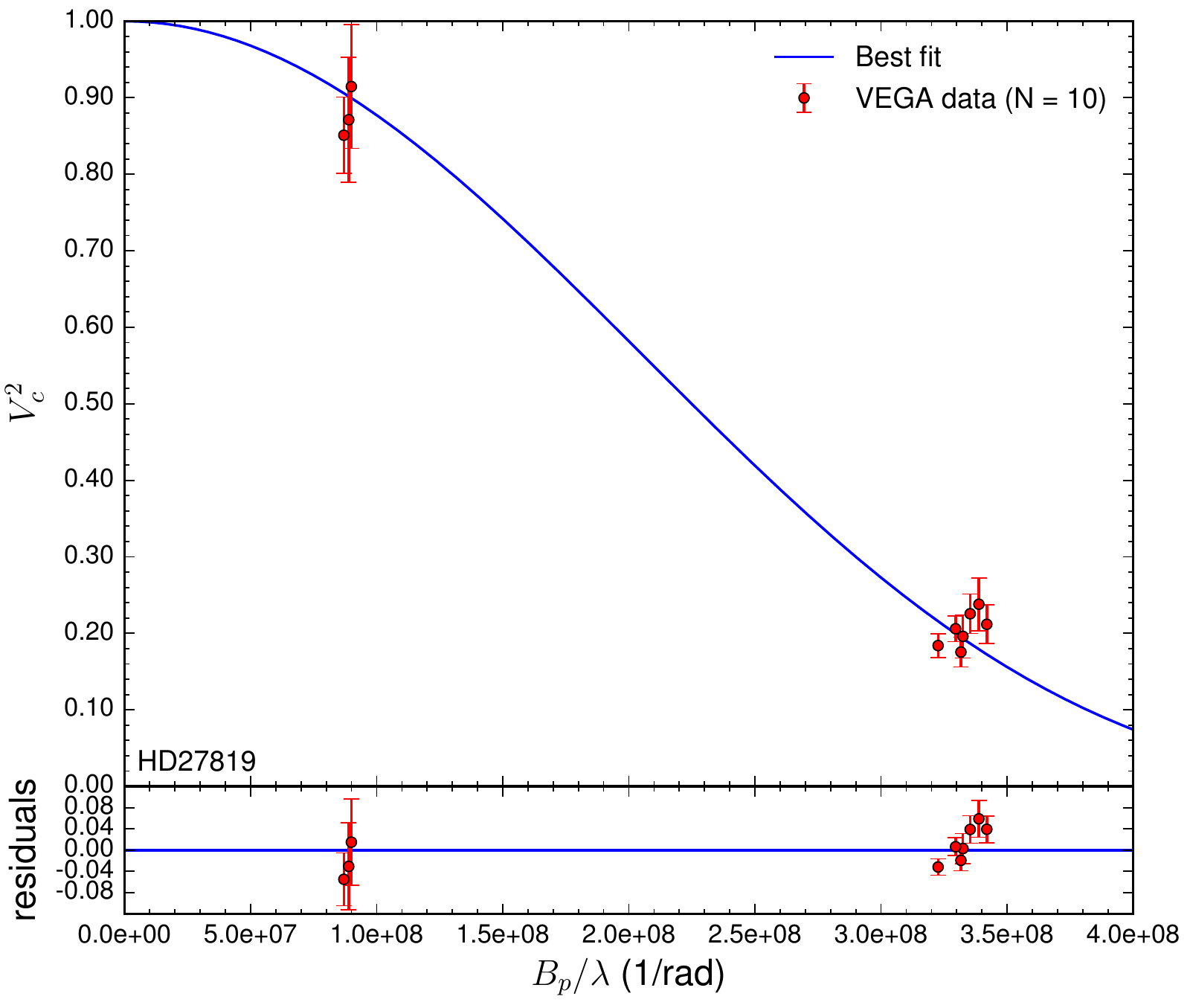}
   \includegraphics[width=0.5\hsize]{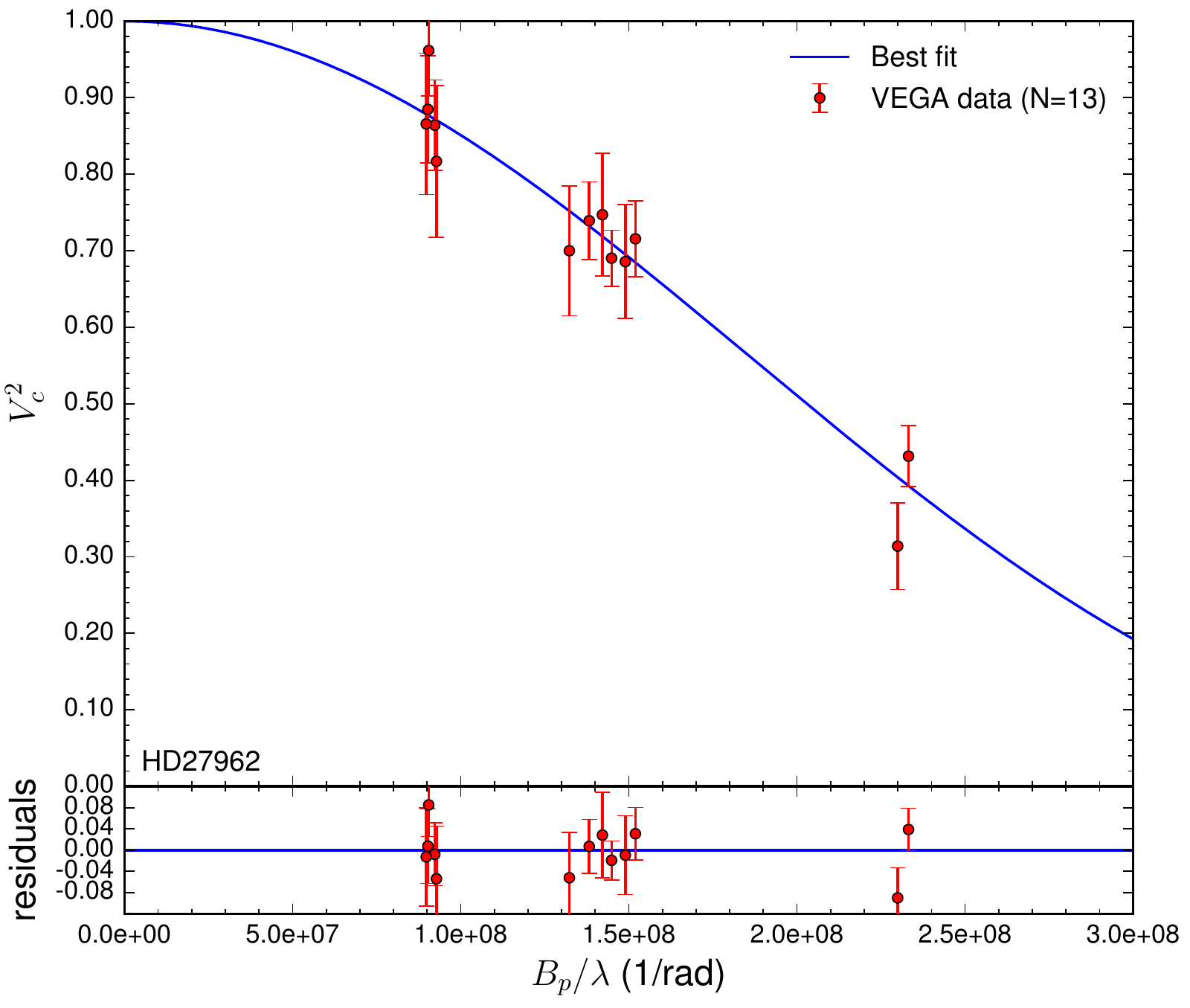}\includegraphics[width=0.5\hsize]{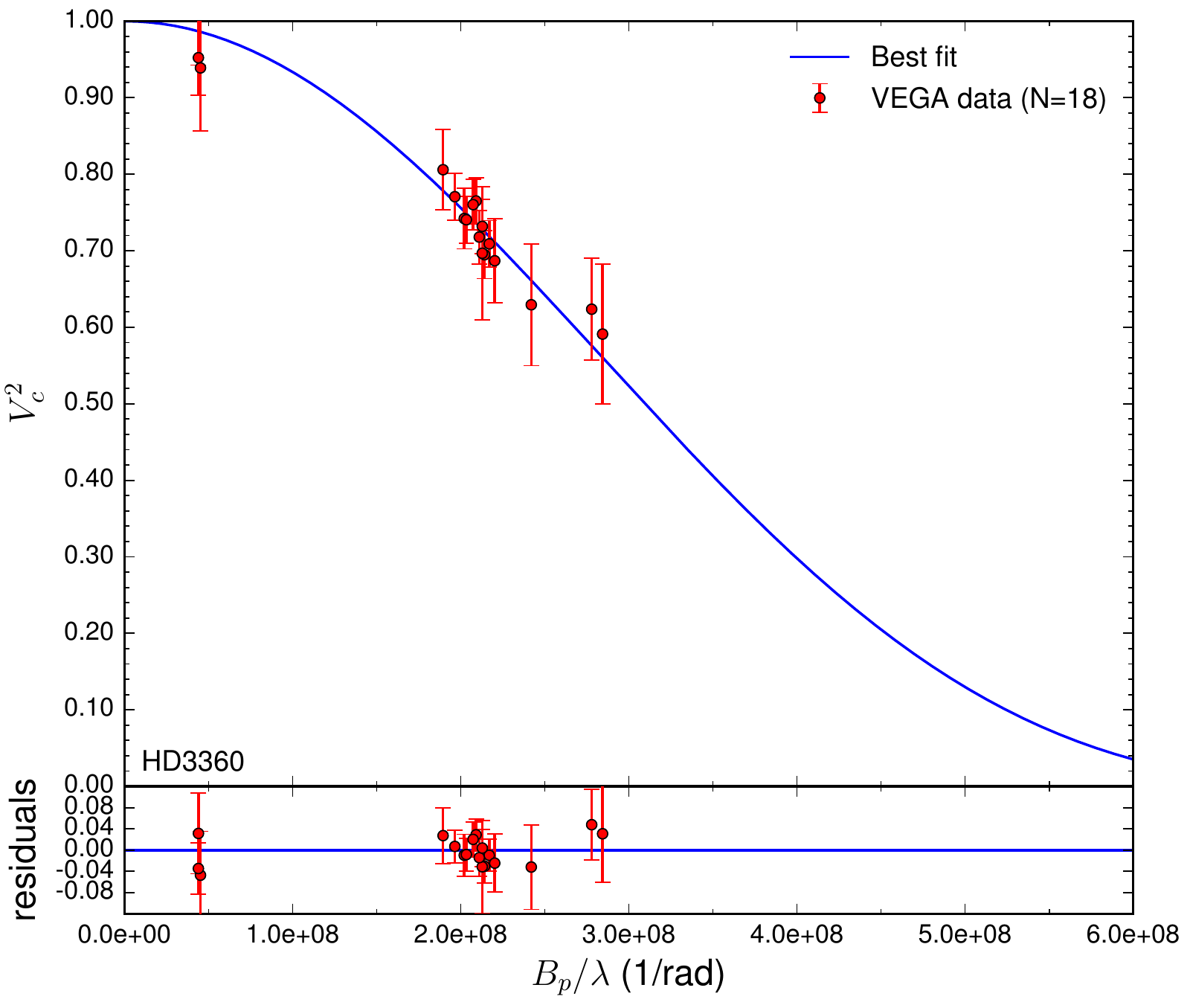}
\end{figure*}

\begin{figure*}  
    \includegraphics[width=0.5\hsize]{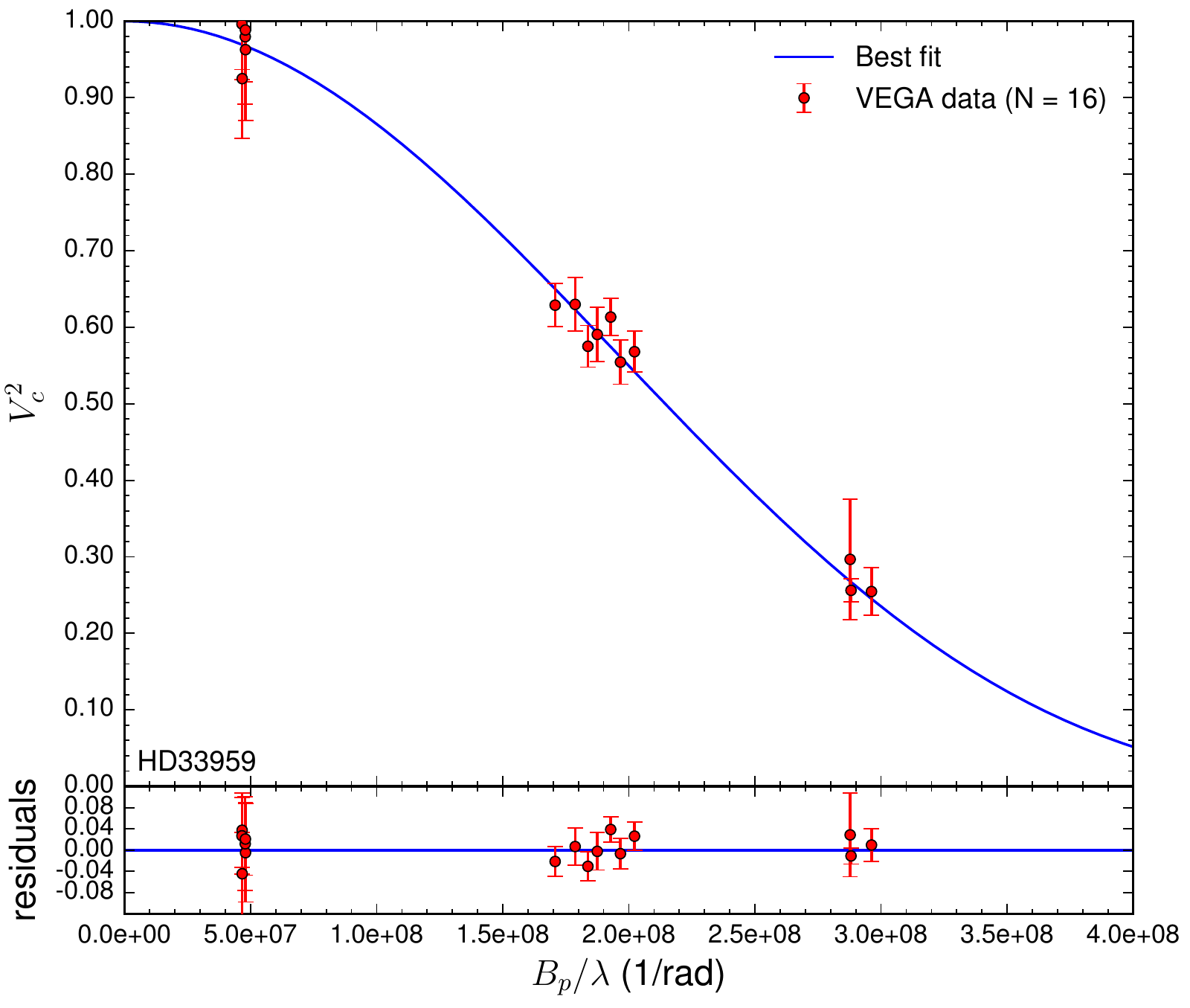}\includegraphics[width=0.5\hsize]{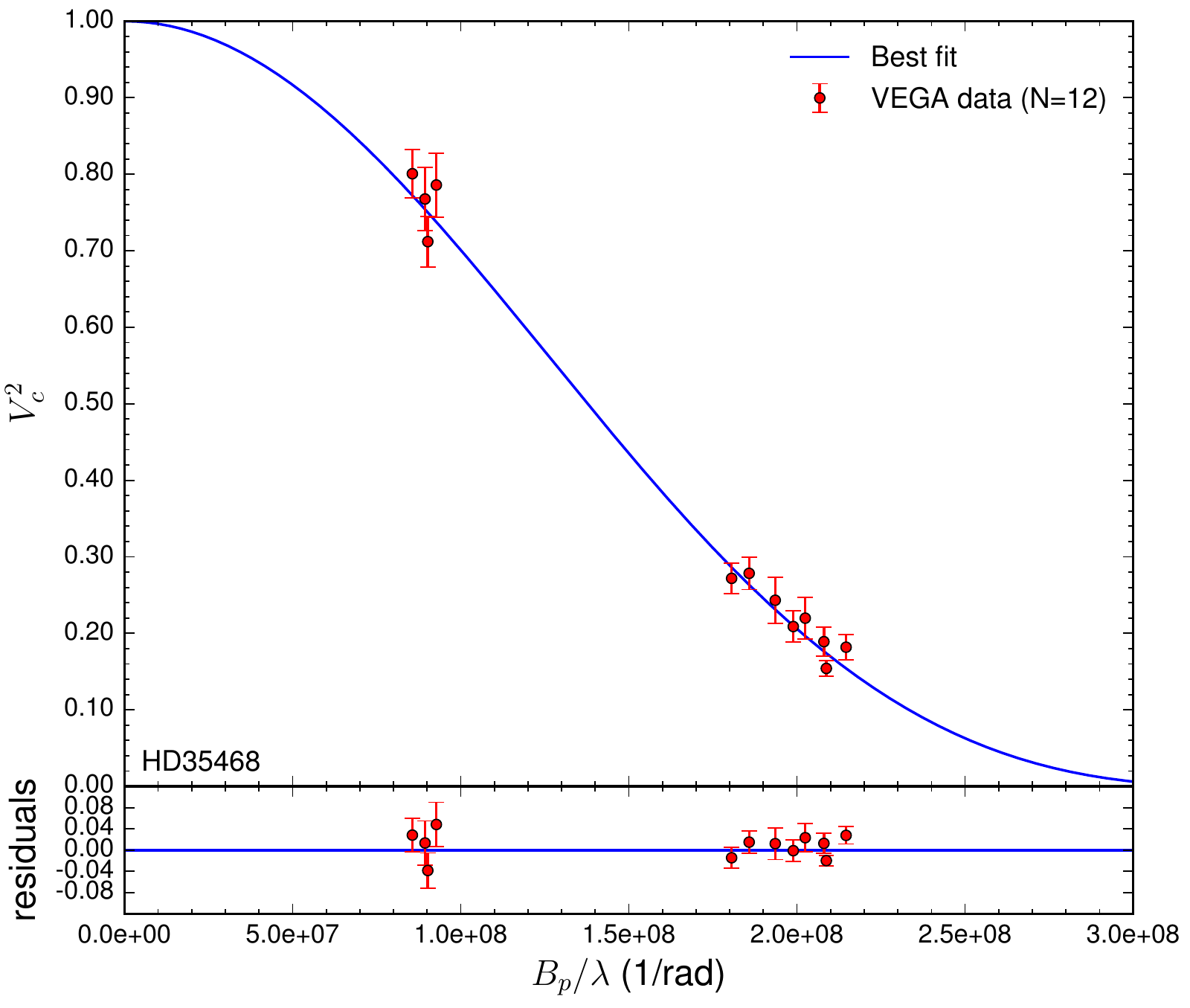}
    \includegraphics[width=0.5\hsize]{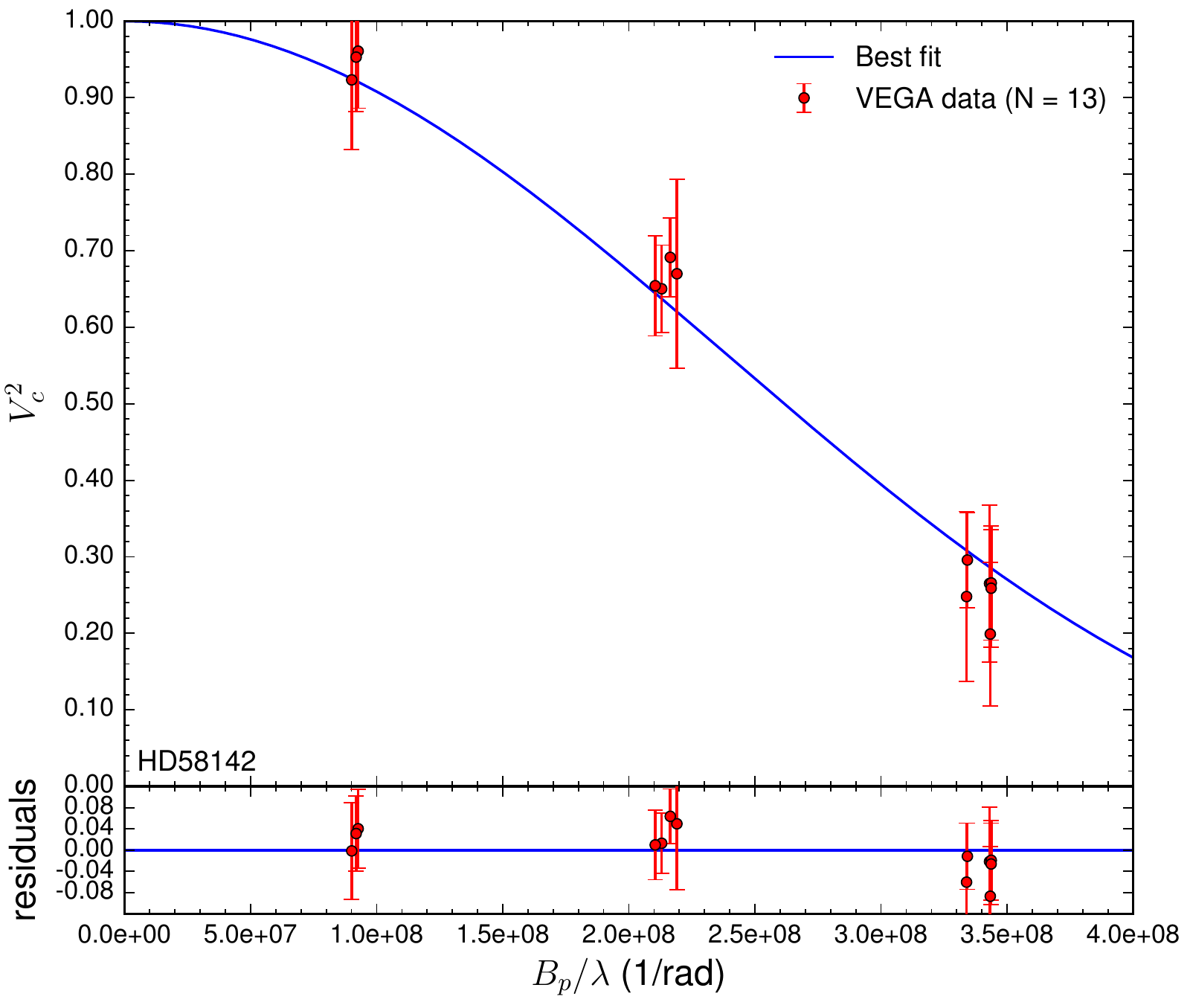}\includegraphics[width=0.5\hsize]{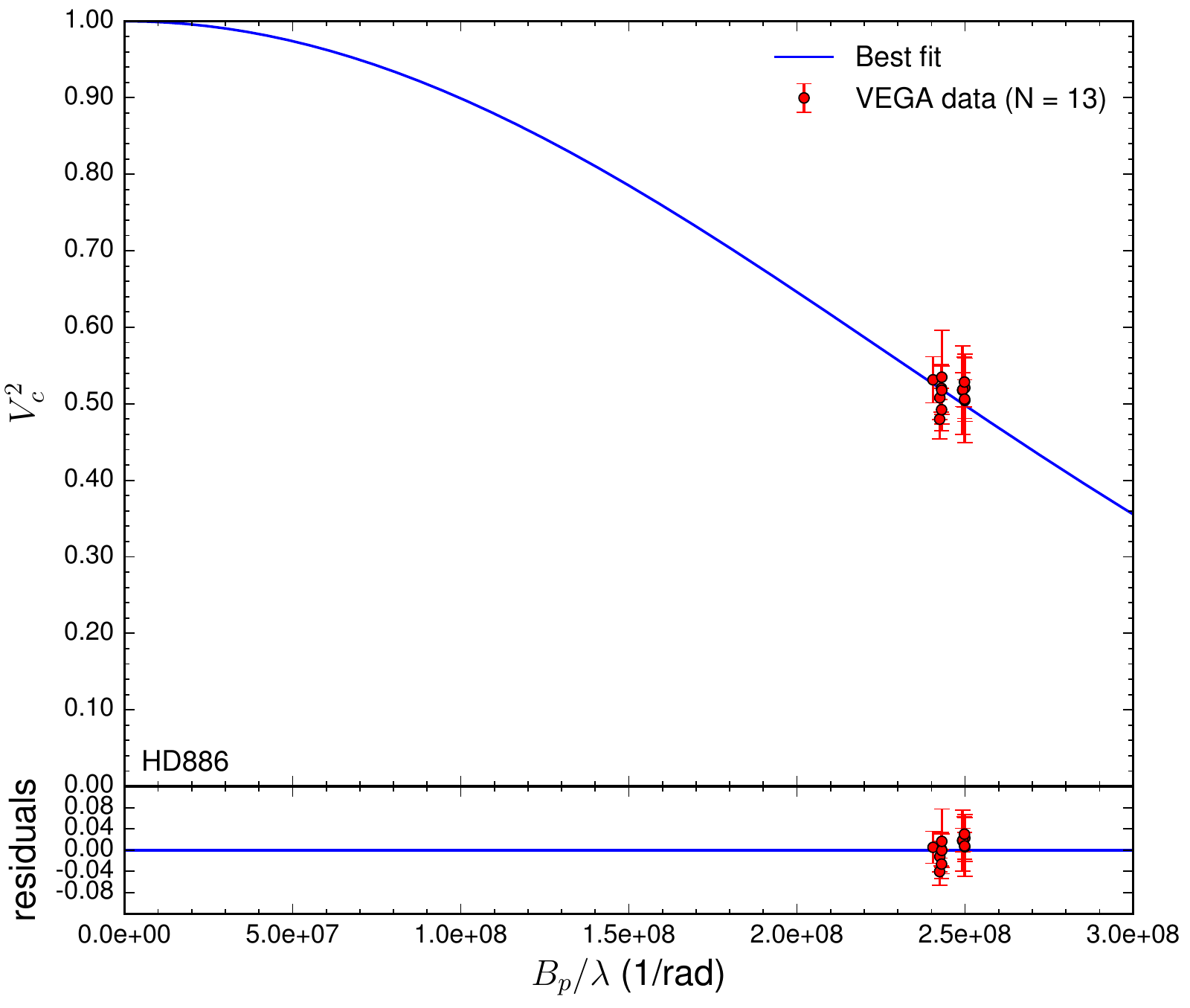}
    \includegraphics[width=0.5\hsize]{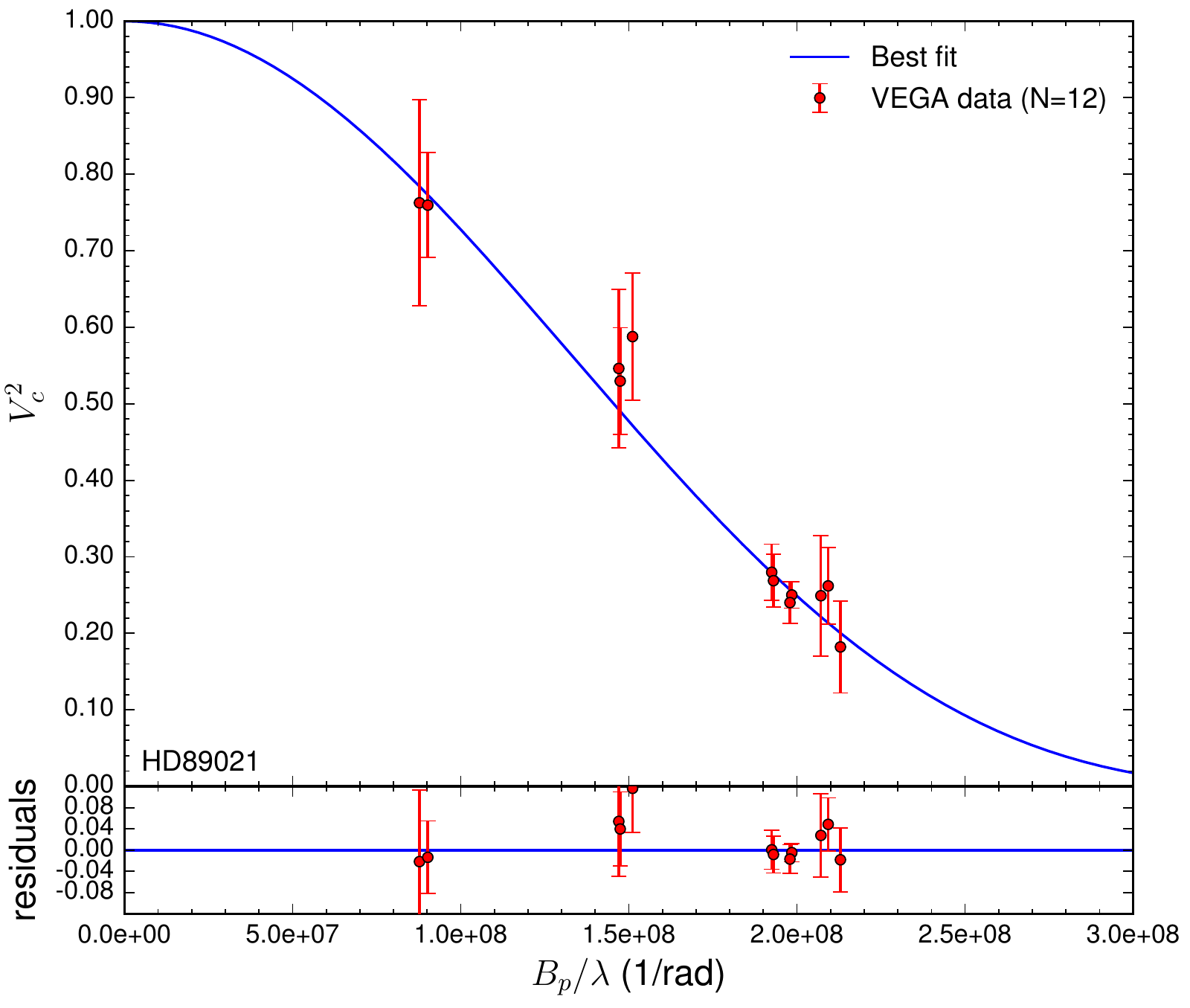}\includegraphics[width=0.5\hsize]{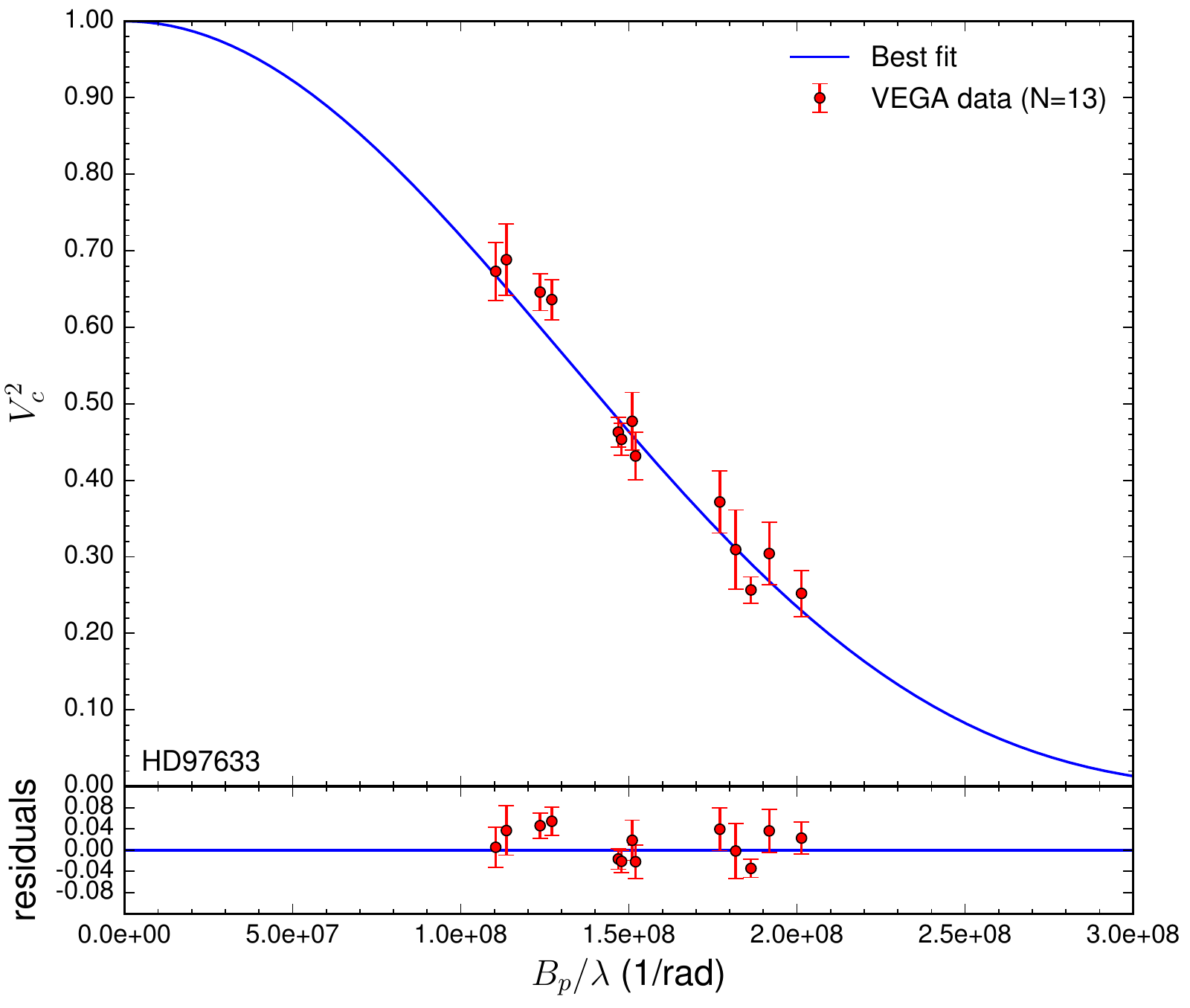}
     \caption{Squared visibility versus spatial frequency for all stars in our sample with their corresponding statistical uncertainties (red dots). The solid blue lines indicate the best uniform-disk model obtained from the LITpro fitting software. See Sect. \ref{VEGA_measurements} for a detailed description of the fitting strategy.}\label{vis_fig}
\end{figure*}

\clearpage
\onecolumn

\section{Observing log}\label{observing_log}

\begin{longtable}{ccccccccccc}
\caption{\label{observing_table} VEGA observing log of the 18 early-type stars. The columns list the date, the hour angle HA, the mean, the minimum, and the maximum wavelengths over which the squared visibility is calculated, the projected baseline length $B_p$, its orientation Arg, and the signal-to-noise ratio (S/N). The last column provides the calibrated squared visibility $V_{\mathrm{cal}}^2$ together with the statistic and the systematic errors. The data are available on the Jean-Marie Mariotti Center OiDB service (available at \url{http://oidb.jmmc.fr}). The interferometric observations are desribed in Sect. \ref{VEGA_measurements}.}\\
\hline
\hline
Star    &       Date    &       Peak    &       HA      &       $\lambda$       &       $\lambda_{\mathrm{min}}$        &       $\lambda_{\mathrm{max}}$        &       $B_p$   &       Arg     &       S/N     &       $V^2_{\mathrm{cal}_{\pm \mathrm{stat} \pm \mathrm{syst}}}$      \\
        &       [yyyy.mm.dd]    &               &       [h]     &       [nm]    &       [nm]    &       [nm]    &       [m]     &       [deg]   &               &               \\
\hline
\endfirsthead
\caption{Continued.}\\
\endhead
\endfoot
\hline                  
HD11415 &       2019.10.07      &       1       &       3.39    &       710     &       700     &       720     &       65.87   &       -165.17 &       15.33   &       0.922   $_{\pm  0.114   \pm     0.003   }$      \\
        &       2019.10.07      &       1       &       3.40    &       730     &       720     &       740     &       65.87   &       -165.27 &       20.17   &       0.898   $_{\pm  0.045   \pm     0.004   }$      \\
        &       2019.10.07      &       1       &       3.79    &       730     &       720     &       740     &       65.87   &       -170.48 &       12.92   &       0.925   $_{\pm  0.072   \pm     0.005   }$      \\
        &       2019.10.07      &       1       &       4.14    &       730     &       720     &       740     &       65.86   &       -175.28 &       18.26   &       0.932   $_{\pm  0.051   \pm     0.004   }$      \\
        &       2019.10.07      &       1       &       4.53    &       730     &       720     &       740     &       65.86   &       179.49  &       11.13   &       0.939   $_{\pm  0.084   \pm     0.004   }$      \\
        &       2020.07.18      &       1       &       -3.41   &       730     &       720     &       740     &       219.83  &       -51.60  &       5.60    &       0.353   $_{\pm  0.063   \pm     0.007   }$      \\
        &       2020.07.18      &       1       &       -2.50   &       730     &       720     &       740     &       229.86  &       -65.39  &       5.26    &       0.286   $_{\pm  0.054   \pm     0.005   }$      \\
        &       2020.07.19      &       1       &       -6.57   &       730     &       720     &       740     &       241.34  &       34.36   &       7.29    &       0.197   $_{\pm  0.027   \pm     0.017   }$      \\
        &       2020.07.19      &       1       &       -5.50   &       730     &       720     &       740     &       244.59  &       21.34   &       5.29    &       0.180   $_{\pm  0.034   \pm     0.015   }$      \\
        &       2020.07.19      &       1       &       -4.98   &       730     &       720     &       740     &       245.45  &       14.99   &       3.33    &       0.225   $_{\pm  0.067   \pm     0.019   }$      \\
        &       2020.07.20      &       1       &       -1.90   &       730     &       720     &       740     &       142.75  &       -89.44  &       11.57   &       0.613   $_{\pm  0.053   \pm     0.015   }$      \\
\hline                                                                                                                                                                                                          
HD114330        &       2019.02.23      &       1       &       0.94    &       730     &       720     &       740     &       182.53  &       12.90   &       4.48    &       0.442   $_{\pm  0.099   \pm     0.025   }$      \\
        &       2019.02.23      &       1       &       2.18    &       710     &       700     &       720     &       177.66  &       -2.94   &       6.54    &       0.535   $_{\pm  0.082   \pm     0.032   }$      \\
        &       2019.02.23      &       1       &       2.21    &       730     &       720     &       740     &       177.73  &       -3.34   &       7.09    &       0.566   $_{\pm  0.080   \pm     0.024   }$      \\
        &       2019.02.23      &       1       &       2.68    &       710     &       700     &       720     &       180.08  &       -9.39   &       7.49    &       0.545   $_{\pm  0.073   \pm     0.033   }$      \\
        &       2019.02.23      &       1       &       2.70    &       730     &       720     &       740     &       180.21  &       -9.61   &       6.96    &       0.522   $_{\pm  0.075   \pm     0.022   }$      \\
        &       2020.03.04      &       1       &       -0.41   &       705     &       695     &       715     &       62.90   &       -114.85 &       17.61   &       0.853   $_{\pm  0.048   \pm     0.005   }$      \\
        &       2020.03.04      &       1       &       -0.41   &       725     &       715     &       735     &       62.90   &       -114.85 &       14.76   &       0.809   $_{\pm  0.055   \pm     0.002   }$      \\
        &       2020.03.04      &       1       &       -0.08   &       705     &       695     &       715     &       61.16   &       -115.11 &       11.59   &       0.829   $_{\pm  0.061   \pm     0.002   }$      \\
        &       2020.03.04      &       1       &       -0.08   &       725     &       715     &       735     &       61.16   &       -115.11 &       9.20    &       0.863   $_{\pm  0.077   \pm     0.002   }$      \\
        &       2020.03.07      &       1       &       -1.26   &       725     &       715     &       735     &       65.54   &       -114.99 &       10.20   &       0.858   $_{\pm  0.084   \pm     0.005   }$      \\
        &       2020.03.07      &       1       &       -0.42   &       705     &       695     &       715     &       62.96   &       -114.84 &       17.71   &       0.903   $_{\pm  0.051   \pm     0.011   }$      \\
        &       2020.03.07      &       1       &       -0.42   &       725     &       715     &       735     &       62.96   &       -114.84 &       9.01    &       0.848   $_{\pm  0.094   \pm     0.002   }$      \\
\hline                                                                                                                                                                                                          
HD145389        &       2019.05.02      &       1       &       2.22    &       730     &       720     &       740     &       151.21  &       -141.88 &       6.65    &       0.586   $_{\pm  0.088   \pm     0.008   }$      \\
        &       2019.05.02      &       1       &       2.69    &       710     &       700     &       720     &       149.07  &       -147.98 &       9.56    &       0.614   $_{\pm  0.094   \pm     0.008   }$      \\
        &       2019.05.02      &       1       &       2.68    &       730     &       720     &       740     &       149.11  &       -147.84 &       11.63   &       0.642   $_{\pm  0.055   \pm     0.010   }$      \\
        &       2019.05.02      &       1       &       2.98    &       710     &       700     &       720     &       147.74  &       -151.96 &       12.80   &       0.590   $_{\pm  0.046   \pm     0.009   }$      \\
        &       2019.05.02      &       1       &       3.00    &       730     &       720     &       740     &       147.66  &       -152.22 &       7.14    &       0.605   $_{\pm  0.085   \pm     0.009   }$      \\
        &       2020.03.04      &       1       &       -1.86   &       725     &       715     &       735     &       61.42   &       -105.17 &       11.66   &       0.914   $_{\pm  0.078   \pm     0.002   }$      \\
\hline                                                                                                                                                                                                          
HD145570        &       2019.06.14      &       1       &       -0.78   &       710     &       700     &       720     &       154.56  &       -110.12 &       9.34    &       0.631   $_{\pm  0.069   \pm     0.013   }$      \\
        &       2019.06.17      &       1       &       -1.96   &       710     &       700     &       720     &       154.79  &       -113.22 &       11.50   &       0.619   $_{\pm  0.054   \pm     0.010   }$      \\
        &       2019.06.17      &       2       &       -1.94   &       710     &       700     &       720     &       83.50   &       87.05   &       5.14    &       0.807   $_{\pm  0.157   \pm     0.006   }$      \\
        &       2019.06.17      &       1       &       -1.62   &       710     &       700     &       720     &       156.03  &       -112.12 &       6.60    &       0.598   $_{\pm  0.091   \pm     0.010   }$      \\
        &       2019.06.17      &       2       &       -1.53   &       710     &       700     &       720     &       90.32   &       88.32   &       13.09   &       0.871   $_{\pm  0.067   \pm     0.004   }$      \\
        &       2019.06.17      &       1       &       -1.22   &       710     &       700     &       720     &       156.14  &       -111.08 &       9.13    &       0.592   $_{\pm  0.065   \pm     0.009   }$      \\
        &       2019.06.17      &       2       &       -1.21   &       710     &       700     &       720     &       94.85   &       89.21   &       8.81    &       0.784   $_{\pm  0.089   \pm     0.004   }$      \\
        &       2019.06.17      &       2       &       -0.86   &       710     &       700     &       720     &       99.11   &       90.15   &       5.86    &       0.858   $_{\pm  0.147   \pm     0.006   }$      \\
\hline                                                                                                                                                                                                          
HD148112        &       2019.06.15      &       1       &       -1.77   &       730     &       720     &       740     &       151.68  &       -108.73 &       8.61    &       0.662   $_{\pm  0.105   \pm     0.033   }$      \\
        &       2019.06.15      &       1       &       -1.20   &       730     &       720     &       740     &       155.38  &       -110.31 &       9.89    &       0.612   $_{\pm  0.062   \pm     0.026   }$      \\
        &       2019.06.15      &       1       &       -0.68   &       730     &       720     &       740     &       156.24  &       -112.09 &       9.80    &       0.647   $_{\pm  0.066   \pm     0.032   }$      \\
        &       2019.08.16      &       1       &       4.37    &       710     &       700     &       720     &       33.77   &       145.58  &       12.78   &       1.002   $_{\pm  0.078   \pm     0.002   }$      \\
        &       2019.08.16      &       1       &       4.39    &       730     &       720     &       740     &       33.76   &       145.54  &       14.82   &       0.995   $_{\pm  0.067   \pm     0.002   }$      \\
\hline                                                                                                                                                                                                          
HD149438        &       2019.06.17      &       1       &       -0.12   &       710     &       700     &       720     &       144.32  &       -102.73 &       13.50   &       0.758   $_{\pm  0.047   \pm     0.010   }$      \\
        &       2019.06.17      &       1       &       0.27    &       710     &       700     &       720     &       138.51  &       -100.46 &       16.78   &       0.791   $_{\pm  0.047   \pm     0.010   }$      \\
        &       2019.06.17      &       1       &       0.28    &       730     &       720     &       740     &       138.35  &       -100.40 &       10.72   &       0.799   $_{\pm  0.075   \pm     0.013   }$      \\
        &       2019.06.17      &       1       &       1.52    &       710     &       700     &       720     &       112.40  &       -92.91  &       15.05   &       0.875   $_{\pm  0.049   \pm     0.011   }$      \\
        &       2019.06.17      &       1       &       1.52    &       730     &       720     &       740     &       112.27  &       -92.88  &       8.79    &       0.816   $_{\pm  0.093   \pm     0.014   }$      \\
        &       2020.03.05      &       1       &       -0.04   &       705     &       695     &       715     &       131.76  &       36.02   &       12.03   &       0.737   $_{\pm  0.082   \pm     0.011   }$      \\
\hline                                                                                                                                                                                                          
HD152107        &       2019.06.15      &       1       &       0.91    &       710     &       700     &       720     &       155.89  &       -126.31 &       8.65    &       0.627   $_{\pm  0.073   \pm     0.013   }$      \\
        &       2019.06.15      &       1       &       0.89    &       730     &       720     &       740     &       155.91  &       -126.155        &       8.33    &       0.614   $_{\pm  0.074   \pm     0.010   }$      \\
        &       2019.06.15      &       1       &       1.53    &       710     &       700     &       720     &       154.36  &       -133.49 &       8.40    &       0.633   $_{\pm  0.075   \pm     0.010   }$      \\
        &       2019.06.15      &       1       &       1.52    &       730     &       720     &       740     &       154.39  &       -133.41 &       11.92   &       0.610   $_{\pm  0.051   \pm     0.009   }$      \\
        &       2020.03.04      &       1       &       -1.09   &       705     &       695     &       715     &       63.97   &       -112.32 &       9.26    &       0.935   $_{\pm  0.101   \pm     0.002   }$      \\
        &       2020.03.04      &       1       &       -1.09   &       725     &       715     &       735     &       63.96   &       -112.30 &       4.81    &       0.873   $_{\pm  0.182   \pm     0.002   }$      \\
        &       2020.03.04      &       1       &       -0.56   &       705     &       695     &       715     &       65.10   &       -117.60 &       12.67   &       0.905   $_{\pm  0.071   \pm     0.003   }$      \\
        &       2020.03.04      &       1       &       -0.56   &       725     &       715     &       735     &       65.09   &       -117.54 &       9.07    &       0.856   $_{\pm  0.094   \pm     0.002   }$      \\
\hline                                                                                                                                                                                                          
HD192640        &       2019.07.06      &       1       &       0.79    &       710     &       700     &       720     &       176.39  &       -27.36  &       5.96    &       0.413   $_{\pm  0.094   \pm     0.005   }$      \\
        &       2019.07.06      &       1       &       0.79    &       730     &       720     &       740     &       176.39  &       -27.38  &       6.88    &       0.427   $_{\pm  0.062   \pm     0.005   }$      \\
        &       2019.07.06      &       1       &       1.17    &       730     &       720     &       740     &       175.53  &       -30.27  &       8.47    &       0.438   $_{\pm  0.052   \pm     0.005   }$      \\
        &       2019.08.15      &       1       &       2.47    &       730     &       720     &       740     &       33.15   &       151.85  &       13.76   &       0.933   $_{\pm  0.052   \pm     0.001   }$      \\
        &       2019.08.15      &       2       &       2.47    &       730     &       720     &       740     &       201.98  &       -37.34  &       5.97    &       0.383   $_{\pm  0.064   \pm     0.006   }$      \\
        &       2019.08.16      &       1       &       1.75    &       710     &       700     &       720     &       33.59   &       156.70  &       17.96   &       1.002   $_{\pm  0.044   \pm     0.001   }$      \\
        &       2019.08.16      &       1       &       1.74    &       730     &       720     &       740     &       33.60   &       156.77  &       11.38   &       0.972   $_{\pm  0.085   \pm     0.001   }$      \\
        &       2019.08.16      &       1       &       2.36    &       710     &       700     &       720     &       33.23   &       152.55  &       23.78   &       0.979   $_{\pm  0.041   \pm     0.001   }$      \\
        &       2019.08.16      &       1       &       2.33    &       730     &       720     &       740     &       33.25   &       152.72  &       19.09   &       0.952   $_{\pm  0.050   \pm     0.001   }$      \\
\hline                                                                                                                                                                                                          
HD195810        &       2019.07.06      &       1       &       1.67    &       710     &       700     &       720     &       174.90  &       -33.63  &       12.65   &       0.571   $_{\pm  0.073   \pm     0.010   }$      \\
        &       2019.07.06      &       1       &       1.66    &       730     &       720     &       740     &       174.86  &       -33.57  &       7.17    &       0.625   $_{\pm  0.087   \pm     0.011   }$      \\
        &       2019.07.06      &       1       &       2.09    &       710     &       700     &       720     &       176.42  &       -35.59  &       11.32   &       0.566   $_{\pm  0.050   \pm     0.011   }$      \\
        &       2019.07.06      &       1       &       2.08    &       730     &       720     &       740     &       176.40  &       -35.56  &       7.55    &       0.580   $_{\pm  0.077   \pm     0.010   }$      \\
        &       2019.07.06      &       1       &       2.43    &       710     &       700     &       720     &       177.17  &       -36.93  &       9.82    &       0.585   $_{\pm  0.060   \pm     0.013   }$      \\
        &       2019.07.06      &       1       &       2.43    &       730     &       720     &       740     &       177.17  &       -36.93  &       8.89    &       0.589   $_{\pm  0.066   \pm     0.011   }$      \\
        &       2019.08.17      &       1       &       2.99    &       710     &       700     &       720     &       33.92   &       149.87  &       23.50   &       0.986   $_{\pm  0.042   \pm     0.001   }$      \\
        &       2019.08.17      &       1       &       3.00    &       730     &       720     &       740     &       33.92   &       149.82  &       19.92   &       0.990   $_{\pm  0.050   \pm     0.001   }$      \\
\hline                                                                                                                                                                                                          
HD27819 &       2019.10.07      &       1       &       -2.12   &       730     &       720     &       740     &       63.50   &       -112.26 &       17.10   &       0.851   $_{\pm  0.050   \pm     0.008   }$      \\
        &       2019.10.07      &       1       &       -1.70   &       730     &       720     &       740     &       64.91   &       -113.64 &       10.64   &       0.871   $_{\pm  0.082   \pm     0.007   }$      \\
        &       2019.10.07      &       1       &       -1.29   &       730     &       720     &       740     &       65.67   &       -115.17 &       11.30   &       0.915   $_{\pm  0.081   \pm     0.008   }$      \\
        &       2019.10.09      &       1       &       -0.89   &       710     &       700     &       720     &       242.78  &       25.72   &       8.39    &       0.212   $_{\pm  0.025   \pm     0.031   }$      \\
        &       2019.10.09      &       1       &       -0.88   &       730     &       720     &       740     &       242.71  &       25.63   &       6.94    &       0.196   $_{\pm  0.028   \pm     0.028   }$      \\
        &       2019.10.09      &       1       &       -0.50   &       710     &       700     &       720     &       240.55  &       22.80   &       6.89    &       0.238   $_{\pm  0.035   \pm     0.033   }$      \\
        &       2019.10.09      &       1       &       -0.51   &       730     &       720     &       740     &       240.58  &       22.84   &       12.26   &       0.206   $_{\pm  0.017   \pm     0.028   }$      \\
        &       2019.10.09      &       1       &       -0.08   &       710     &       700     &       720     &       238.10  &       19.40   &       8.07    &       0.226   $_{\pm  0.026   \pm     0.028   }$      \\
        &       2019.10.09      &       1       &       0.39    &       710     &       700     &       720     &       235.49  &       15.25   &       8.39    &       0.176   $_{\pm  0.020   \pm     0.022   }$      \\
        &       2019.10.09      &       1       &       0.38    &       730     &       720     &       740     &       235.52  &       15.31   &       11.87   &       0.184   $_{\pm  0.015   \pm     0.024   }$      \\
\hline                                                                                                                                                                                                          
HD27962 &       2019.10.07      &       1       &       -0.90   &       710     &       700     &       720     &       65.87   &       -116.84 &       8.26    &       0.817   $_{\pm  0.099   \pm     0.010   }$      \\
        &       2019.10.07      &       1       &       -0.92   &       730     &       720     &       740     &       65.87   &       -116.77 &       12.64   &       0.885   $_{\pm  0.070   \pm     0.009   }$      \\
        &       2019.10.07      &       1       &       -0.45   &       710     &       700     &       720     &       65.52   &       -119.09 &       14.64   &       0.864   $_{\pm  0.059   \pm     0.009   }$      \\
        &       2019.10.07      &       1       &       -0.45   &       730     &       720     &       740     &       65.51   &       -119.11 &       9.38    &       0.866   $_{\pm  0.092   \pm     0.007   }$      \\
        &       2019.10.07      &       1       &       0.10    &       710     &       700     &       720     &       64.24   &       -122.33 &       16.21   &       0.962   $_{\pm  0.059   \pm     0.010   }$      \\
        &       2019.10.10      &       1       &       -1.67   &       710     &       700     &       720     &       165.56  &       -6.20   &       1.81    &       0.432   $_{\pm  0.040   \pm     0.019   }$      \\
        &       2019.10.10      &       1       &       -1.67   &       720     &       700     &       740     &       165.56  &       -6.17   &       5.53    &       0.314   $_{\pm  0.057   \pm     0.019   }$      \\
        &       2020.12.16      &       1       &       -1.51   &       710     &       700     &       720     &       93.93   &       105.62  &       8.24    &       0.700   $_{\pm  0.085   \pm     0.023   }$      \\
        &       2020.12.16      &       1       &       -0.88   &       710     &       700     &       720     &       100.92  &       101.70  &       9.30    &       0.747   $_{\pm  0.080   \pm     0.017   }$      \\
        &       2020.12.16      &       1       &       -0.89   &       730     &       720     &       740     &       100.89  &       101.72  &       14.54   &       0.739   $_{\pm  0.051   \pm     0.021   }$      \\
        &       2020.12.16      &       1       &       -0.22   &       710     &       700     &       720     &       105.76  &       98.17   &       9.23    &       0.686   $_{\pm  0.074   \pm     0.016   }$      \\
        &       2020.12.16      &       1       &       -0.23   &       730     &       720     &       740     &       105.74  &       98.19   &       18.87   &       0.690   $_{\pm  0.037   \pm     0.020   }$      \\
        &       2020.12.16      &       1       &       0.50    &       710     &       700     &       720     &       107.88  &       94.72   &       14.49   &       0.716   $_{\pm  0.049   \pm     0.020   }$      \\
\hline                                                                                                                                                                                                          
HD3360  &       2019.08.15      &       1       &       -0.68   &       710     &       700     &       720     &       32.08   &       175.73  &       11.37   &       0.939   $_{\pm  0.083   \pm     0.001   }$      \\
        &       2019.08.15      &       2       &       -0.71   &       710     &       700     &       720     &       201.90  &       -13.57  &       6.21    &       0.591   $_{\pm  0.091   \pm     0.007   }$      \\
        &       2019.08.15      &       1       &       -0.70   &       730     &       720     &       740     &       32.09   &       175.85  &       19.57   &       0.952   $_{\pm  0.049   \pm     0.001   }$      \\
        &       2019.08.16      &       1       &       -1.26   &       730     &       720     &       740     &       32.11   &       -179.09 &       13.46   &       1.018   $_{\pm  0.076   \pm     0.001   }$      \\
        &       2019.08.16      &       2       &       -1.26   &       730     &       720     &       740     &       202.87  &       -8.25   &       9.36    &       0.624   $_{\pm  0.067   \pm     0.011   }$      \\
        &       2020.08.26      &       1       &       -2.75   &       705     &       695     &       715     &       133.59  &       -83.32  &       15.32   &       0.806   $_{\pm  0.053   \pm     0.005   }$      \\
        &       2020.08.26      &       1       &       -1.96   &       705     &       695     &       715     &       142.47  &       -93.06  &       18.79   &       0.742   $_{\pm  0.039   \pm     0.010   }$      \\
        &       2020.08.26      &       1       &       -1.96   &       725     &       715     &       735     &       142.47  &       -93.06  &       24.95   &       0.771   $_{\pm  0.031   \pm     0.010   }$      \\
        &       2020.08.26      &       1       &       -1.40   &       705     &       695     &       715     &       147.43  &       -99.61  &       25.44   &       0.765   $_{\pm  0.030   \pm     0.015   }$      \\
        &       2020.08.26      &       1       &       -1.40   &       725     &       715     &       735     &       147.43  &       -99.61  &       24.16   &       0.741   $_{\pm  0.031   \pm     0.013   }$      \\
        &       2020.08.26      &       1       &       -1.00   &       725     &       715     &       735     &       150.34  &       -104.32 &       22.90   &       0.760   $_{\pm  0.033   \pm     0.014   }$      \\
        &       2020.08.26      &       1       &       0.13    &       705     &       695     &       715     &       155.25  &       -117.31 &       12.55   &       0.687   $_{\pm  0.055   \pm     0.012   }$      \\
        &       2020.08.36      &       1       &       0.13    &       725     &       715     &       735     &       155.25  &       -117.31 &       22.27   &       0.695   $_{\pm  0.031   \pm     0.009   }$      \\
        &       2020.08.29      &       1       &       -1.94   &       710     &       700     &       720     &       171.79  &       -3.35   &       7.91    &       0.629   $_{\pm  0.080   \pm     0.015   }$      \\
        &       2020.08.30      &       1       &       -0.87   &       710     &       700     &       720     &       151.08  &       -105.72 &       7.99    &       0.697   $_{\pm  0.087   \pm     0.011   }$      \\
        &       2020.08.30      &       1       &       -0.26   &       710     &       700     &       720     &       154.04  &       -112.78 &       23.36   &       0.709   $_{\pm  0.030   \pm     0.015   }$      \\
        &       2020.08.30      &       1       &       -0.26   &       730     &       720     &       740     &       154.04  &       -112.78 &       20.52   &       0.718   $_{\pm  0.035   \pm     0.012   }$      \\
        &       2020.08.30      &       1       &       0.18    &       730     &       720     &       740     &       155.36  &       -117.88 &       20.73   &       0.732   $_{\pm  0.035   \pm     0.010   }$      \\
\hline                                                                                                                                                                                                          
HD33959 &       2019.08.17      &       1       &       -3.64   &       710     &       700     &       720     &       34.00   &       -159.96 &       11.11   &       0.980   $_{\pm  0.088   \pm     0.001   }$      \\
        &       2019.08.17      &       1       &       -3.65   &       730     &       720     &       740     &       34.00   &       -159.95 &       13.75   &       0.996   $_{\pm  0.072   \pm     0.001   }$      \\
        &       2019.08.17      &       1       &       -3.26   &       710     &       700     &       720     &       34.05   &       -162.80 &       14.53   &       0.989   $_{\pm  0.068   \pm     0.001   }$      \\
        &       2019.08.17      &       2       &       -3.24   &       710     &       700     &       720     &       210.32  &       10.48   &       8.18    &       0.255   $_{\pm  0.031   \pm     0.004   }$      \\
        &       2019.08.17      &       1       &       -3.27   &       730     &       720     &       740     &       34.05   &       -162.76 &       14.37   &       1.007   $_{\pm  0.070   \pm     0.001   }$      \\
        &       2019.08.17      &       2       &       -3.24   &       730     &       720     &       740     &       210.31  &       10.42   &       16.78   &       0.256   $_{\pm  0.015   \pm     0.004   }$      \\
        &       2019.08.17      &       1       &       -2.81   &       710     &       700     &       720     &       34.07   &       -166.35 &       10.39   &       0.963   $_{\pm  0.093   \pm     0.001   }$      \\
        &       2019.08.17      &       1       &       -2.83   &       730     &       720     &       740     &       34.07   &       -166.15 &       11.85   &       0.925   $_{\pm  0.078   \pm     0.001   }$      \\
        &       2019.08.17      &       2       &       -2.82   &       730     &       720     &       740     &       210.06  &       6.55    &       3.78    &       0.297   $_{\pm  0.079   \pm     0.005   }$      \\
        &       2020.08.26      &       1       &       -3.41   &       705     &       695     &       715     &       120.39  &       -91.44  &       22.24   &       0.629   $_{\pm  0.028   \pm     0.056   }$      \\
        &       2020.08.26      &       1       &       -2.97   &       705     &       695     &       715     &       129.56  &       -94.82  &       21.21   &       0.575   $_{\pm  0.027   \pm     0.084   }$      \\
        &       2020.08.26      &       1       &       -2.97   &       725     &       715     &       735     &       129.56  &       -94.82  &       18.09   &       0.630   $_{\pm  0.035   \pm     0.072   }$      \\
        &       2020.08.26      &       1       &       -2.61   &       705     &       695     &       715     &       135.92  &       -97.39  &       25.38   &       0.613   $_{\pm  0.024   \pm     0.071   }$      \\
        &       2020.08.26      &       1       &       -2.61   &       725     &       715     &       735     &       135.92  &       -97.39  &       16.74   &       0.591   $_{\pm  0.035   \pm     0.005   }$      \\
        &       2020.08.26      &       1       &       -2.18   &       705     &       695     &       715     &       142.58  &       -100.48 &       21.21   &       0.568   $_{\pm  0.027   \pm     0.083   }$      \\
        &       2020.08.26      &       1       &       -2.18   &       725     &       715     &       735     &       142.58  &       -100.48 &       19.17   &       0.554   $_{\pm  0.029   \pm     0.068   }$      \\
\hline                                                                                                                                                                                                          
HD35468 &       2019.10.05      &       1       &       -1.20   &       710     &       700     &       720     &       65.85   &       -115.71 &       18.80   &       0.786   $_{\pm  0.042   \pm     0.004   }$      \\
        &       2019.10.05      &       1       &       -1.19   &       730     &       720     &       740     &       65.85   &       -115.72 &       21.47   &       0.712   $_{\pm  0.033   \pm     0.003   }$      \\
        &       2019.10.05      &       1       &       -0.73   &       730     &       720     &       740     &       65.27   &       -116.74 &       18.46   &       0.768   $_{\pm  0.042   \pm     0.003   }$      \\
        &       2019.10.05      &       1       &       0.08    &       730     &       720     &       740     &       62.49   &       -119.41 &       25.37   &       0.801   $_{\pm  0.032   \pm     0.003   }$      \\
        &       2020.08.30      &       1       &       -3.28   &       710     &       700     &       720     &       131.90  &       -111.05 &       13.14   &       0.278   $_{\pm  0.021   \pm     0.017   }$      \\
        &       2020.08.30      &       1       &       -3.29   &       730     &       720     &       740     &       131.81  &       -111.06 &       13.84   &       0.272   $_{\pm  0.020   \pm     0.017   }$      \\
        &       2020.08.30      &       1       &       -2.77   &       710     &       700     &       720     &       141.22  &       -110.42 &       10.25   &       0.209   $_{\pm  0.020   \pm     0.020   }$      \\
        &       2020.08.30      &       1       &       -2.77   &       730     &       720     &       740     &       141.28  &       -110.41 &       8.12    &       0.243   $_{\pm  0.030   \pm     0.004   }$      \\
        &       2020.08.30      &       1       &       -2.33   &       710     &       700     &       720     &       147.69  &       -110.21 &       10.09   &       0.189   $_{\pm  0.019   \pm     0.018   }$      \\
        &       2020.08.30      &       1       &       -2.32   &       730     &       720     &       740     &       147.80  &       -110.21 &       8.17    &       0.220   $_{\pm  0.027   \pm     0.004   }$      \\
        &       2020.08.30      &       1       &       -1.88   &       710     &       700     &       720     &       152.37  &       -110.29 &       11.20   &       0.182   $_{\pm  0.016   \pm     0.017   }$      \\
        &       2020.08.30      &       1       &       -1.88   &       730     &       720     &       740     &       152.37  &       -110.29 &       15.35   &       0.154   $_{\pm  0.010   \pm     0.003   }$      \\
\hline                                                                                                                                                                                                          
HD58142 &       2019.02.23      &       1       &       1.07    &       710     &       700     &       720     &       243.55  &       8.38    &       2.58    &       0.265   $_{\pm  0.103   \pm     0.006   }$      \\
        &       2019.02.23      &       1       &       1.66    &       710     &       700     &       720     &       244.00  &       2.83    &       3.37    &       0.259   $_{\pm  0.077   \pm     0.008   }$      \\
        &       2019.02.23      &       1       &       1.66    &       730     &       720     &       740     &       244.00  &       2.77    &       4.75    &       0.296   $_{\pm  0.062   \pm     0.011   }$      \\
        &       2019.02.23      &       1       &       2.14    &       710     &       700     &       720     &       244.03  &       -1.76   &       3.56    &       0.266   $_{\pm  0.075   \pm     0.009   }$      \\
        &       2019.02.23      &       1       &       2.63    &       710     &       700     &       720     &       243.76  &       -6.49   &       2.12    &       0.199   $_{\pm  0.094   \pm     0.007   }$      \\
        &       2019.02.23      &       1       &       2.62    &       730     &       720     &       740     &       243.77  &       -6.34   &       3.50    &       0.248   $_{\pm  0.111   \pm     0.012   }$      \\
        &       2020.03.07      &       1       &       1.83    &       705     &       695     &       715     &       65.319  &       -144.30 &       12.88   &       0.961   $_{\pm  0.075   \pm     0.008   }$      \\
        &       2020.03.07      &       2       &       1.83    &       705     &       695     &       715     &       154.43  &       -137.53 &       5.42    &       0.670   $_{\pm  0.124   \pm     0.037   }$      \\
        &       2020.03.07      &       1       &       1.83    &       725     &       715     &       735     &       65.322  &       -144.25 &       10.16   &       0.923   $_{\pm  0.091   \pm     0.004   }$      \\
        &       2020.03.07      &       2       &       1.83    &       725     &       715     &       735     &       154.43  &       -137.53 &       11.41   &       0.650   $_{\pm  0.057   \pm     0.037   }$      \\
        &       2020.03.07      &       1       &       2.43    &       705     &       695     &       715     &       64.76   &       -151.88 &       13.43   &       0.953   $_{\pm  0.071   \pm     0.008   }$      \\
        &       2020.03.07      &       2       &       2.43    &       705     &       695     &       715     &       152.58  &       -145.25 &       13.37   &       0.691   $_{\pm  0.052   \pm     0.037   }$      \\
        &       2020.03.07      &       2       &       2.43    &       725     &       715     &       735     &       152.58  &       -145.25 &       10.00   &       0.654   $_{\pm  0.065   \pm     0.044   }$      \\
\hline                                                                                                                                                                                                          
HD886   &       2019.08.20      &       1       &       1.42    &       730     &       720     &       740     &       175.54  &       -31.95  &       17.64   &       0.531   $_{\pm  0.030   \pm     0.005   }$      \\
        &       2019.08.20      &       1       &       1.95    &       710     &       700     &       720     &       176.97  &       -34.74  &       8.98    &       0.518   $_{\pm  0.058   \pm     0.005   }$      \\
        &       2019.08.20      &       1       &       1.95    &       730     &       720     &       740     &       176.97  &       -34.74  &       18.83   &       0.480   $_{\pm  0.025   \pm     0.004   }$      \\
        &       2019.08.20      &       1       &       2.49    &       710     &       700     &       720     &       177.44  &       -37.09  &       9.13    &       0.504   $_{\pm  0.055   \pm     0.005   }$      \\
        &       2019.08.20      &       1       &       2.49    &       730     &       720     &       740     &       177.44  &       -37.09  &       16.47   &       0.518   $_{\pm  0.031   \pm     0.005   }$      \\
        &       2019.08.21      &       1       &       2.48    &       710     &       700     &       720     &       177.44  &       -37.05  &       11.85   &       0.521   $_{\pm  0.044   \pm     0.005   }$      \\
        &       2019.08.21      &       1       &       2.48    &       730     &       720     &       740     &       177.44  &       -37.05  &       8.78    &       0.535   $_{\pm  0.061   \pm     0.005   }$      \\
        &       2019.08.22      &       1       &       1.97    &       710     &       700     &       720     &       177.02  &       -34.86  &       23.36   &       0.519   $_{\pm  0.022   \pm     0.005   }$      \\
        &       2019.08.22      &       1       &       1.97    &       730     &       720     &       740     &       177.02  &       -34.86  &       17.84   &       0.508   $_{\pm  0.028   \pm     0.005   }$      \\
        &       2019.08.22      &       1       &       2.28    &       710     &       700     &       720     &       177.39  &       -36.26  &       20.02   &       0.506   $_{\pm  0.025   \pm     0.005   }$      \\
        &       2019.08.22      &       1       &       2.28    &       730     &       720     &       740     &       177.39  &       -36.24  &       17.81   &       0.492   $_{\pm  0.028   \pm     0.004   }$      \\
        &       2019.08.22      &       1       &       2.63    &       710     &       700     &       720     &       177.37  &       -37.62  &       16.20   &       0.528   $_{\pm  0.033   \pm     0.005   }$      \\
        &       2019.08.22      &       1       &       2.63    &       730     &       720     &       740     &       177.37  &       -37.62  &       16.68   &       0.520   $_{\pm  0.031   \pm     0.005   }$      \\
\hline                                                                                                                                                                                                          
HD89021 &       2019.02.26      &       1       &       1.99    &       710     &       700     &       720     &       64.05   &       -145.57 &       8.43    &       0.760   $_{\pm  0.069   \pm     0.007   }$      \\
        &       2019.02.26      &       2       &       2.00    &       710     &       700     &       720     &       151.20  &       -138.74 &       7.89    &       0.182   $_{\pm  0.060   \pm     0.007   }$      \\
        &       2019.02.26      &       1       &       2.00    &       730     &       720     &       740     &       64.04   &       -145.63 &       5.66    &       0.763   $_{\pm  0.135   \pm     0.006   }$      \\
        &       2019.02.26      &       2       &       2.00    &       730     &       720     &       740     &       151.20  &       -138.74 &       3.17    &       0.249   $_{\pm  0.079   \pm     0.010   }$      \\
        &       2019.02.26      &       2       &       2.50    &       710     &       700     &       720     &       148.56  &       -145.21 &       7.36    &       0.262   $_{\pm  0.050   \pm     0.007   }$      \\
        &       2020.06.15      &       1       &       4.88    &       710     &       700     &       720     &       140.52  &       179.97  &       8.83    &       0.240   $_{\pm  0.027   \pm     0.002   }$      \\
        &       2020.06.15      &       1       &       4.89    &       730     &       720     &       740     &       140.52  &       179.75  &       7.65    &       0.280   $_{\pm  0.037   \pm     0.003   }$      \\
        &       2020.06.15      &       1       &       5.35    &       710     &       700     &       720     &       140.91  &       172.67  &       8.54    &       0.250   $_{\pm  0.017   \pm     0.002   }$      \\
        &       2020.06.15      &       1       &       5.34    &       730     &       720     &       740     &       140.90  &       172.78  &       7.76    &       0.269   $_{\pm  0.035   \pm     0.002   }$      \\
        &       2020.12.19      &       1       &       -0.15   &       710     &       700     &       720     &       107.28  &       101.47  &       7.04    &       0.588   $_{\pm  0.083   \pm     0.045   }$      \\
        &       2020.12.20      &       1       &       -0.14   &       730     &       720     &       740     &       107.31  &       101.33  &       5.28    &       0.546   $_{\pm  0.104   \pm     0.020   }$      \\
        &       2020.12.21      &       1       &       0.80    &       730     &       720     &       740     &       107.62  &       91.73   &       7.59    &       0.530   $_{\pm  0.070   \pm     0.021   }$      \\
\hline                                                                                                                                                                                                          
HD97633 &       2019.05.03      &       1       &       1.13    &       710     &       700     &       720     &       142.96  &       -122.06 &       8.38    &       0.252   $_{\pm  0.030   \pm     0.007   }$      \\
        &       2019.05.03      &       1       &       1.59    &       710     &       700     &       720     &       136.12  &       -125.81 &       8.83    &       0.304   $_{\pm  0.041   \pm     0.009   }$      \\
        &       2019.05.03      &       1       &       1.60    &       730     &       720     &       740     &       136.02  &       -125.87 &       14.95   &       0.257   $_{\pm  0.017   \pm     0.006   }$      \\
        &       2019.05.03      &       1       &       2.01    &       710     &       700     &       720     &       129.31  &       -129.75 &       8.94    &       0.309   $_{\pm  0.052   \pm     0.013   }$      \\
        &       2019.05.03      &       1       &       2.01    &       730     &       720     &       740     &       129.25  &       -129.78 &       9.18    &       0.372   $_{\pm  0.040   \pm     0.013   }$      \\
        &       2020.12.16      &       1       &       -2.35   &       710     &       700     &       720     &       80.64   &       110.16  &       14.83   &       0.688   $_{\pm  0.046   \pm     0.031   }$      \\
        &       2020.12.16      &       1       &       -2.35   &       730     &       720     &       740     &       80.59   &       110.18  &       17.82   &       0.673   $_{\pm  0.038   \pm     0.029   }$      \\
        &       2020.12.16      &       1       &       -1.74   &       710     &       700     &       720     &       90.23   &       105.65  &       24.23   &       0.636   $_{\pm  0.026   \pm     0.029   }$      \\
        &       2020.12.16      &       1       &       -1.74   &       730     &       720     &       740     &       90.23   &       105.65  &       26.70   &       0.646   $_{\pm  0.024   \pm     0.028   }$      \\
        &       2020.12.16      &       1       &       0.14    &       710     &       700     &       720     &       107.21  &       96.11   &       12.62   &       0.477   $_{\pm  0.038   \pm     0.022   }$      \\
        &       2020.12.16      &       1       &       0.14    &       730     &       720     &       740     &       107.21  &       96.11   &       23.73   &       0.463   $_{\pm  0.020   \pm     0.020   }$      \\
        &       2020.12.16      &       1       &       0.61    &       710     &       700     &       720     &       107.92  &       94.19   &       13.77   &       0.432   $_{\pm  0.031   \pm     0.019   }$      \\
        &       2020.12.16      &       1       &       0.62    &       730     &       720     &       740     &       107.91  &       94.18   &       21.60   &       0.454   $_{\pm  0.021   \pm     0.019   }$      \\
\hline                                                                                                                                                                                                                                                                  
\end{longtable}    

\end{appendix}

\end{document}